\documentclass[3p,11pt]{elsarticle}

\usepackage{amssymb,amsmath,amsfonts,latexsym}
\usepackage{graphicx,epstopdf}
\usepackage{subfigure}
\newtheorem{definition}{Definition}
\newcommand{\Id}{1\!\!1}

\makeatletter
\def\ps@pprintTitle{%
   \let\@oddhead\@empty
   \let\@evenhead\@empty
   \def\@oddfoot{\reset@font\hfil\thepage\hfil}
   \let\@evenfoot\@oddfoot
}
\makeatother

\begin{document}

\begin{frontmatter}

\title{$(H,\rho)$--induced dynamics and large time behaviors}

\author[l1,l3,l4]{F. Bagarello}
\ead{fabio.bagarello@unipa.it}

\author[l2]{R. Di Salvo}
\ead{rosa.disalvo@unime.it}

\author[l1]{F. Gargano}
\ead{francesco.gargano@unipa.it}

\author[l2]{F. Oliveri\corref{cor}}
\ead{francesco.oliveri@unime.it}
\cortext[cor]{Corresponding author}

\address[l1]{DEIM --  University of Palermo,
Viale delle Scienze, I--90128  Palermo, Italy}
\address[l2]{Department MIFT -- University of Messina,
Viale F. Stagno d'Alcontres 31, I--98166 Messina, Italy}
\address[l3]{I.N.F.N -  Sezione di Napoli}
\address[l4]{Department of Mathematics and Applied Mathematics,
University of Cape Town, South Africa}

\begin{abstract}
In some recent papers, the so called  $(H,\rho)$-induced dynamics of a system $\mathcal{S}$ whose time evolution is deduced adopting an operatorial approach, borrowed in part from quantum mechanics, has been introduced. Here, $H$ is the Hamiltonian for $\mathcal{S}$, while $\rho$ is a certain rule applied periodically (or not) on $\mathcal{S}$. The analysis carried on throughout this paper  shows that, replacing the Heisenberg
dynamics with the $(H,\rho)$-induced one, we obtain a simple, and somehow natural, way to prove that some relevant dynamical variables of $\mathcal{S}$ may converge, for large $t$, to certain asymptotic values. This can not be so, for finite dimensional systems, if no rule is considered. In this case, in fact, any Heisenberg dynamics implemented by a suitable hermitian operator $H$ can only give an oscillating behavior. We prove our claims both analytically and numerically for a simple system with two degrees of freedom, and then we apply our general scheme to a model describing a biological system of bacteria living in a two-dimensional lattice, where two different choices of the rule
are considered.\end{abstract}


\end{frontmatter}

\section{Introduction}
\label{sec:intro}

Since 2006 \cite{bag2006}, it has become clear that raising and lowering operators of quantum mechanics can be  successfully used for the mathematical description of some macroscopic systems.
Many applications are discussed in \cite{bagbook}, and others, not necessarily related to ladder operators but still to quantum ideas, have been recently considered
in order to model the dynamics in several contexts, such as social life and decision-making processes \cite{qdm1,qdm3,qdm4,qdm5,Pkren1,BHK,rev3c}, population and crowd dynamics \cite{BO_migration,BGO_crowds,GAR14,BTBB}, ecological processes \cite{BO_ecomod,BCO_desert,DSO_RM2016,DSO_AAPP2016},  political systems \cite{pol1,pol4,all1,all2,all3,BG17,DSO_turncoat2017,DSGO_opinion2017,DSGO_turncoat2017,rev3b}.

The key aspect of the operatorial approach adopted here is that the time evolution of an observable $X$ of the macroscopic system $\mathcal{S}$ we are considering is given by $X(t)=\exp(iHt)X\exp(-iHt)$, where $H$ is the time independent self-adjoint Hamiltonian of $\mathcal{S}$, and  the mean values of the observable $X$ are linked to real valued functions phenomenologically associated to some macroscopic quantities. This approach has revealed efficient in predict real dynamics.

Not surprisingly, the description of the dynamics of operatorial models ruled by a self-adjoint time independent Hamiltonian has some limitations. For instance, all the observables of a system $\mathcal{S}$ with a finite number of degrees of freedom and governed by a quadratic Hamiltonian operator, exhibit, at most, a quasiperiodic dynamics. Then, if we wish to model a system  $\mathcal{S}$ having an asymptotic \emph{final state}, it is clear that such a description does not work, and the modeling framework needs to be enriched, if not completely changed.
This is not new: in quantum optics, or for two or three level atoms, if we need to describe a transition from one level to another, in some cases  some effective finite dimensional non-hermitian Hamiltonians are used. In this way, decays are well described phenomenologically. Another way consists in considering the atoms interacting with some infinite reservoir, but  in such a case the full system
($\mathcal{S}_{full}$, \emph{i.e.}, $\mathcal{S}$ plus the reservoir) is no longer finite-dimensional. Sometimes one describes a simplified version of $\mathcal{S}_{full}$ by means of some master equation where, again, the dynamics is described by some finite matrix whose entries are properly chosen; \emph{per contra},  many details of the system are lost with this kind of description, and the physics could be somehow hidden.

In  \cite{BDSGO_GoL}, where a quantum version of game of life has been considered, we have proposed an extended version of the Heisenberg dynamics in order to take into account effects which may occur during the time evolution of the system $\mathcal{S}$, and which can not apparently be included in a purely Hamiltonian description. In particular, at fixed times of the evolution of a system $\mathcal{S}$ driven by a hermitian  time independent Hamiltonian $H$, some checks on $\mathcal{S}$ are performed, and  used to change the state of the system itself according to some explicit prescription.  
A slightly different viewpoint has been proposed in \cite{DSO_AAPP2016,DSO_turncoat2017,DSGO_opinion2017,DSGO_turncoat2017}, where the periodical checks on the state of the system are used to change some of the parameters entering the Hamiltonian, without modifying the functional form of the Hamiltonian itself. This approach proved to be quite efficient in operatorial models of stressed bacterial populations \cite{DSO_AAPP2016}, as well as in models of political parties affected by turncoat--like behaviors \cite{DSO_turncoat2017,DSGO_opinion2017,DSGO_turncoat2017}.

Here, we want to show that this $(H,\rho)$--induced dynamics can be efficiently used to describe (finite-dimensional) systems going to some equilibrium, even when $H$ is hermitian. This strategy may give interesting results if the rule $\rho$ is not introduced as a mere mathematical trick, but  is somehow physically justified. For instance,  in \cite{DSO_AAPP2016}, the rule accounts for the modifications in the metabolic activity of bacteria due to lack of nutrients and/or to the presence of waste material, whereas, in \cite{DSO_turncoat2017,DSGO_opinion2017,DSGO_turncoat2017}, the rule modifies the behavior of the members of a political party with regard to their tendency to shift allegiance from one loyalty or ideal to another one.

The paper is organized as follows. In Section~\ref{sect2}, we briefly review the main definitions and results concerning the two possible strategies for  $(H,\rho)$--induced dynamics. In Section~\ref{sect3}, we describe a very simple two-mode system, and discuss the role of the rule $\rho$ in the emergence of an asymptotic time evolution.  Section~\ref{sect4} is devoted to the analysis of a more complicated system, \emph{i.e.}, a model for a bacterial population on a two-dimensional lattice.
Section~\ref{sect5} contains our conclusions and perspectives. Finally, in the appendix, we consider a situation where 
a simple time dependent Hamiltonian may produce a dynamics with an asymptotic state, at least under very special choices of its parameters.

\section{Extending the dynamics with rules: $(H,\rho)$-induced dynamics}
\label{sect2}

To keep the paper self-contained, in this Section we briefly discuss the mathematical framework of the $(H,\rho)$--induced dynamics (further details can be found in \cite{DSO_AAPP2016,DSO_turncoat2017,DSGO_opinion2017,DSGO_turncoat2017,BDSGO_GoL}). 
The proposed approaches merge the general framework of quantum dynamics, described by a Hamiltonian $H$, with some periodic (or not) effects which can not be included in $H$.

Let $\mathcal{S}$ be our physical system, and $Q_j$ ($j=1,\ldots,M$) a set of $M$ commuting self-adjoint operators with eigenvectors $\varphi^{(j)}_{\alpha_n}$ and eigenvalues $\alpha_n^{(j)}$:
\begin{equation}
[Q_j,Q_k]=Q_jQ_k-Q_kQ_j=0, \qquad Q_j=Q_j^\dagger,\qquad  Q_j\varphi^{(j)}_{n_j}=\alpha_{n_j}^{(j)}\varphi^{(j)}_{n_{j}},
\label{21}
\end{equation}
$j,k=1,2,\ldots,M$, $n_j=1,2,3,\ldots,N_j$. Setting $\mathbf{n}=(n_1,n_2,\ldots,n_M)$, the vector
\begin{equation}
\varphi_{\mathbf{n}}=\varphi^{(1)}_{n_{1}}\otimes\varphi^{(2)}_{n_{2}}\otimes\cdots\varphi^{(M)}_{n_{M}}
\end{equation}
represents an eigenstate of all the operators $Q_j$, say
\begin{equation}
Q_j\,\varphi_{\mathbf{n}}=\alpha_{n_j}^{(j)}\,\varphi_{\mathbf{n}}.
\label{22}
\end{equation}
The existence of a common eigenstate for all the operators $Q_j$ is guaranteed by the fact that they all mutually commute. It is convenient, and always true in our applications, to assume that these vectors are mutually orthogonal and normalized:
\begin{equation}
\left<\varphi_{\mathbf{n}},\varphi_{\mathbf{m}}\right>=\delta_{\mathbf{n},\mathbf{m}}=\prod_{j=1}^M\delta_{n_j,m_j}.
\label{23}
\end{equation}
The Hilbert space $\mathcal{H}$ where $\mathcal{S}$ is  defined is mathematically constructed as the closure of the linear span of all the vectors $\varphi_\mathbf{n}$,  defining an orthonormal basis for $\mathcal{H}$. Now, let $H$ be a quadratic
time independent self-adjoint Hamiltonian, describing the interactions occurring in $\mathcal{S}$, plus its \emph{free kinetic contribution}. Notice that $H$, in general, does not commute with the $Q_j$'s. In absence of any other information, the wave function
$\Psi(t)$, describing $\mathcal{S}$ at time $t$, evolves according to the
Schr\"odinger equation $i\dot\Psi(t)=H\Psi(t)$, where $\Psi(0)=\Psi_0$ describes the initial state of $\mathcal{S}$.
The formal solution of the Schr\"odinger equation is, since $H$ does not depend explicitly on $t$, $\Psi(t)=\exp(-iHt)\Psi(0)=\exp(-iHt)\Psi_0$.
We can now compute the mean value of each operator $Q_j$ in the state $\Psi(t)$,
\begin{equation}
q_j(t)=\left<\Psi(t),Q_j\Psi(t)\right>,
\end{equation}
and use it to define a related $M$-dimensional time dependent vector $\mathbf{q}(t)=(q_1(t),\allowbreak q_2(t),\allowbreak \ldots,\allowbreak q_M(t))$.

It is well known \cite{Merzbacher,Messiah} that this is not the unique way to look at the time evolution of $\mathcal{S}$. Another equivalent way consists in adopting the Heisenberg representation, in which the wave function does not evolve in time, while the operators do, according to the equation $\dot X(t)=i[H,X(t)]$, whereupon we have $X(t)=\exp(iHt)X(0)\exp(-iHt)$. Here, $X(t)$ is a generic operator of $\mathcal{S}$ acting on $\mathcal{H}$ at time $t$. In this approach, if $X$ is an observable of the system, after fixing a vector state
$\varphi_\mathbf{n}$ representing the initial configuration of the system, we may compute its mean value
\begin{equation}
x(t)=\left<\varphi_\mathbf{n},X(t)\varphi_\mathbf{n}\right>,
\end{equation}
that, in the case of fermionic operatorial models, if $X$ is the number operator, can be interpreted as a  local density \cite{bagbook}.

We are now ready to introduce, rather generally, two different ways of extending the dynamics through the introduction of a specific \emph{rule}.

In the first approach, the rule $\rho$ is a  map from $\mathcal{H}$ to $\mathcal{H}$.
Its explicit action depends on the expression of $\mathbf{q}(t)$ at particular instants $k\tau$ ($k\in\mathbb{N}, \tau\in\mathbb{R}^+$). In other words, according to how $\mathbf{q}(k\tau)$ looks like, $\rho$ maps the input vector $\Psi(\tau)$ (here $k=1$) into a different output vector $\Psi_{new}$, and we write  $\rho(\Psi(\tau))=\Psi_{new}$.
This is not very different from what happens in scattering theory, where an incoming state, after the occurrence of the scattering, is transformed into an outgoing state \cite{roman}.  The new vector $\Psi_{new}$ can be considered as the new initial state of the system driven, for another time interval of length $\tau$, by the Hamiltonian $H$, and this procedure can be iterated several times. As an example, in \cite{BDSGO_GoL}, the rule $\rho$ is used to map an incoming state into one of the elements of the basis of $\mathcal{H}$, allowing to extend the classical concept of rule used in a cellular automaton. Our approach has some aspects in common with the \textit{repeated quantum measurements}, according to which the state of a system is perturbed by a non trivial quantum measurement, and subsequent measurements of the same system reveal the presence of disturbances if compared to the situation in which no previous measurements were performed. In our case, the disturbance manifests with the creation of the new state  $\Psi_{new}$. The key difference in our approach lies in the fact that
the new generated state is prepared as the result of the choice of the rules, so that it is the result of a \textit{controlled post-processing} phase. On the other hand, in the classical notion of repeated quantum measurements, the new state is the result of disturbances that may not be controlled depending on the kind of apparatus used for the measurements \cite{HHK2016}. Here, we also like to mention that the repeated action of the rule $\rho$ in our approach is similar to what is done in the context of the quantum Zeno effect (see \cite{pascazio}, and references therein), where some measures on a quantum system are repeated again and again.

In the second approach, the rule $\rho$ works on the space of the parameters of the Hamiltonian $H$, rather than on the wave function. In this case, given a quadratic time independent
self-adjoint Hamiltonian involving $p$ real parameters, the rule $\rho$ is a map 
from $\mathbb{R}^p$ to $\mathbb{R}^p$
that, at particular instants $k\tau$, on the basis of the actual state of the system (or of its variation), changes
some of the values of these parameters. In such a way, in some sense, the model \emph{adjusts} itself.

\subsection{The rule $\rho$ as a map from $\mathcal{H}$ to $\mathcal{H}$}
\label{subsec:rhoHil}
Let us briefly sketch some general considerations about the rule $\rho$  as a map from $\mathcal{H}$ to $\mathcal{H}$.
First, let us observe that there exists a one-to-one correspondence between $\mathbf{n}$ and the vector $\varphi_\mathbf{n}$:
once we know $\mathbf{n}$, $\varphi_\mathbf{n}$ is clearly identified, and  vice-versa. Suppose now that at
time $t=0$ the system $\mathcal{S}$ is in a state $\mathbf{n}^0$ or, which is the same, $\mathcal{S}$ is described by the vector
$\varphi_{\mathbf{n}^0}$. Then, once fixed a positive value of $\tau$, this vector evolves
 in the time interval $[0,\tau[$ according to the Schr\"odinger recipe: $\exp(-iHt)\varphi_{\mathbf{n}^0}$. Let us set
\begin{equation}
 \Psi(\tau^-)=\lim_{t\rightarrow \tau^-}\exp(-iHt)\varphi_{\mathbf{n}^0},
\end{equation}
where $t$ converges to $\tau$ from below\footnote{We use here $\tau^-$, $2\tau^-$, $\ldots$,  as argument of $\Psi$ to emphasize that {\bf before} $\tau$, for instance, the time evolution is only due to $H$, while $\rho$ really acts at $t=\tau$.}. Now, at time $t=\tau$, $\rho$ is applied to $\Psi(\tau^-)$, and the output of this action is a
new vector which we assume here to be again an eigenstate of each operator $Q_j$, but with different eigenvalues, $\varphi_{\mathbf{n}^1}$\footnote{This choice is not the only possibility to set up a rule. In fact, other possibilities can also be considered. The key common point to all possible choices is that $\rho$ behaves as a check over the system $\mathcal{S}$, and modifies some of its ingredients according to the result of this check. See also Section \ref{subsec:rhopar}.}. In other words, $\rho$ \emph{looks} at the  explicit expression of  $\Psi(\tau^-)$ and, according to its form, returns a new vector $\mathbf{n}^1=(n^1_1,n^1_2,\ldots,n^1_{M})$;  as a consequence, a
new vector $\varphi_{\mathbf{n}^1}$ of $\mathcal{H}$ is obtained. Examples of how $\rho$ explicitly acts are discussed in Section \ref{sect3}. Now, the procedure is iterated, taking $\varphi_{\mathbf{n}^1}$
as the initial vector, and letting it evolve with $H$ for another time interval of length $\tau$;
we compute
\begin{equation}
\Psi(2\tau^-)=\lim_{t\rightarrow \tau^-}\exp(-iHt)\varphi_{\mathbf{n}^1},
\end{equation}
and the new vector  $\varphi_{\mathbf{n}^2}$ is deduced by the action of rule $\rho$ on $\Psi(2\tau^-)$: $\varphi_{\mathbf{n}^2}=\rho(\Psi(2\tau^-))$.   Then, in general, for all $k\geq1$, we have
\begin{equation}
\Psi(k\tau^-)=\lim_{t\rightarrow \tau^-}\exp(-iHt)\varphi_{\mathbf{n}^{k-1}},\qquad
\varphi_{\mathbf{n}^k}=\rho \left(\Psi(k\tau^-)\right).
\label{add2}
\end{equation}

Now, let $X$ be a generic operator on $\mathcal{H}$, bounded or unbounded. In this latter case, we will require that the various $\varphi_{\mathbf{n}^k}$  belong to the domain of $X(t)=\exp(iHt)X\exp(-iHt)$ for all $t\in[0,\tau]$.

\begin{definition}\label{def1}
The sequence of functions
\begin{equation}
x_{k+1}(t):=\left<\varphi_{\mathbf{n}^k}, X(t)\varphi_{\mathbf{n}^k}\right>,
\label{24}
\end{equation}
for $t\in[0,\tau]$ and $k\in {\mathbb{N}}_0$, is called the $(H,\rho)$--induced dynamics of $X$.
\end{definition}

Some consequences of Definition~\ref{def1} and some properties of the sequence ${\underline X}(\tau)=(x_1(\tau),x_2(\tau),\ldots)$ have been discussed in \cite{BDSGO_GoL}. Moreover,
 from ${\underline X}(t)=(x_1(t),\allowbreak x_2(t),\ldots)$ it is possible to define a new function of time, giving the evolution of the observable $X$ of $\mathcal{S}$, in the following way:
\begin{equation}
\tilde X(t)=
\left\{
\begin{array}{ll}
x_1(t)\qquad &t\in [0,\tau[  , \\
x_2(t-\tau)\qquad &t\in [\tau,2\tau[  , \\
x_3(t-2\tau)\qquad &t\in [2\tau,3\tau[ ,  \\
\ldots &
\end{array}
\right.
\label{25bis}
\end{equation}
It is clear that $\tilde X(t)$ may have discontinuities in $k\tau$, for positive
integers $k$.

\subsection{The rule $\rho$ as a map in the space of the parameters of $H$}
\label{subsec:rhopar}
In the previous subsection, the effect of $\rho$ was to change the state of the system, from an input to an output vector. The other \emph{elements} of $\mathcal{S}$, in particular its Hamiltonian, stay unchanged at each step. We now discuss a different approach where a rule acts on $\mathcal{S}$ changing some aspects of the dynamics of $\mathcal{S}$ related to the Hamiltonian operator.  

Let $\mathcal{S}$ be a system involving $M$ fermionic (or bosonic) modes and suppose that its evolution is ruled by the following quadratic time independent self-adjoint Hamiltonian
\begin{equation}
\label{Ham}
H=\sum_{j=1}^M\omega_j a_j^\dagger a_j+\sum_{j=1}^{M-1}\sum_{k=j+1}^M\lambda_{j,k}(a_ja_k^\dagger+a_ka_j^\dagger),
\end{equation}
involving the $p=M(M+1)/2$  real parameters (not necessarily all non vanishing) $\omega_j$ and $\lambda_{j,k}$, where $a_j$ and $a_j^\dagger$, $i=1,\ldots M$, are annihilation and creation operators, respectively.

Adopting the Heisenberg representation, the time evolution of the lowering operators $a_j$'s is given by
\begin{equation}
a_j(t)=\exp(iHt)a_j(0)\exp(-iHt),\qquad j=1,\ldots,M,
\end{equation}
or,
equivalently, by the solution of the following linear system of ordinary differential equations:
\begin{equation}
\label{eqs_Heis}
\dot a_j(t)=i\left(-\omega_j a_j(t)+\sum_{k=1,k\neq j}^M \lambda_{j,k}a_k(t)\right),
\qquad j=1,\ldots,M.
\end{equation}

Restricting ourselves to the fermionic case, in principle, we have a system of $M 2^{2M}$ linear differential equations to be solved\footnote{For bosons the situation is, in general, more tricky since we have to deal with unbounded operators in an infinite-dimensional Hilbert space.} with suitable initial conditions for the matrices representing the operators $a_j$. However, since the system is linear, we may write it in compact form, say
\begin{equation}
\dot A(t)=U A(t),
\label{NA1}
\end{equation}
where $A(t)=\left(a_1(t),  a_2(t), \ldots, a_M(t)\right)^T$,
and $U$ is an $M\times M$ constant matrix such that $U_{j,j}=-i\omega_j$,
$U_{j,k}=i\lambda_{j,k}$, and each component of $A$ is a $2^M\times 2^M$  matrix. The formal solution is immediately deduced, namely
\begin{equation}
A(t)=\exp(U t)A(0)=V(t)A(0).
\label{NA2}
\end{equation}
Thus, if $V_{\ell,m}(t)$ is the generic entry of matrix $V(t)$, we have
\begin{equation}
a_\ell(t)=\sum_{k=1}^M V_{\ell,k}(t)a_k(0), \qquad \ell=1,\ldots,M.
\end{equation}

Now, we need to compute the mean value of the number operator for the $\ell$-th mode (which is intended to represent a physical quantity which is relevant for the description of $\mathcal{S}$)
\begin{equation}
\hat n_\ell(t)=a_\ell^\dagger(t)a_\ell(t)
\end{equation}
on an eigenvector $\varphi_{n_1,n_2,\ldots,n_M}$ of all the $\hat n_\ell(0)$,
\begin{equation}
\hat n_\ell\varphi_{n_1,n_2,\ldots,n_M}=n_\ell\varphi_{n_1,n_2,\ldots,n_M}, \qquad \ell=1,2,\ldots,M.
\end{equation}
It is easy to check that the relations
\begin{equation}
n_\ell(t)=\left<\varphi_{n_1,n_2,\ldots,n_M},\hat n_\ell(t)\varphi_{n_1,n_2,\ldots,n_M}\right>=
\sum_{k=1}^M \left|V_{\ell,k}(t)\right|^2n_k,
\label{NA3}
\end{equation}
where $\ell=1,\ldots,M$, provide what we are looking for.

Starting from a quadratic Hamiltonian such as the one defined in (\ref{Ham}), we may enrich the dynamics by introducing a rule, repeatedly acting at specific instants, and accounting for a sort of dependence of the parameters $\omega_j$ and $\lambda_{j,k}$ in (\ref{Ham}) upon the current state of the system.   In some sense, the model \emph{adjusts} itself as a consequence of its evolution. We stress that in such an approach only the strengths of the mutual interactions change, whereas the model preserves its functional structure. 

Here is a sketch of this approach. Let us start considering a self-adjoint quadratic Hamiltonian operator $H^{(1)}$, the corresponding evolution of a certain observable $X$
\begin{equation}
X(t)=\exp(iH^{(1)}t)X\exp(-iH^{(1)}t),
\end{equation}
and compute its mean value
\begin{equation}\label{add3}
x(t)=\langle\varphi_{n_1,n_2,\ldots,n_M},\,
X(t)\varphi_{n_1,n_2,\ldots,n_M}\rangle
\end{equation}
in a time interval of length $\tau>0$ on a vector $\varphi_{n_1,n_2,\ldots,n_M}$. Then, let us modify some of the parameters involved in $H^{(1)}$, on the basis of the values of the various $x(\tau)$ according to (\ref{add3}). In this way, we get a new Hamiltonian operator $H^{(2)}$, having the same functional form as $H^{(1)}$, but (in general) with  different values of (some of) the involved parameters, and follow the continuous evolution
of the system\footnote{The evolution of $\mathcal{S}$ is no more stopped and restarted from a modified state, as done in \ref{subsec:rhoHil}, but rather the initial vector state for each subinterval corresponds exactly to the state reached by the system.} under the action of this new Hamiltonian for the next time interval of length $\tau$. 
Actually, we do not restart the evolution of the system from a new initial condition, but simply continue to follow the evolution with the only difference that for $t\in]\tau,2\tau]$ a new Hamiltonian $H^{(2)}$ rules the process. 
And so on. Therefore, the rule now has to be thought of as a map from $\mathbb{R}^p$ into $\mathbb{R}^p$ acting on the space of the parameters involved in the Hamiltonian. Therefore, the global evolution is governed by a sequence of similar Hamiltonian operators, and the parameters entering the model can be considered stepwise (in time) constant.

In general terms, let us consider a time interval $[0,T]$, and split it in $n=T/\tau$ subintervals of length $\tau$. Assume  $n$ to be integer. In the $k$-th subinterval $[(k-1)\tau,k\tau[$ consider an Hermitian Hamiltonian $H^{(k)}$ ruling the dynamics. The global dynamics arises from the sequence of Hamiltonians
\begin{equation}
H^{(1)} \stackrel{\tau}{\longrightarrow} H^{(2)} \stackrel{\tau}{\longrightarrow} H^{(3)} \stackrel{\tau}{\longrightarrow} \ldots \stackrel{\tau}{\longrightarrow} H^{(n)},
\end{equation}
the complete evolution being obtained by glueing the local evolutions.

In every subinterval we therefore have a system like
\begin{equation}
\label{eq:sub_sys}
\dot A(t)=U^{(k)}A(t),  \qquad t \in [(k-1)\tau,k\tau].
\end{equation}

To obtain the mean values of the number operators at each instant $t$, we need the computation of the exponential of the $M\times M$ matrix $U^{(k)} t$.
There are various methods in the literature, based on different grounds (see \cite{Moler}, and the references therein, for a survey on this topic), for carrying out the exponential of large matrices.
The best general algorithms use matrix decomposition methods;
they start with the Schur decomposition and include some sort of eigenvalue
clustering. There are also variants which involve further reduction to a block form. In all
cases the initial decomposition costs $O(M^3)$. Therefore, roughly speaking, the computational complexity
for the exponential of a matrix of order  $M$ can be considered as $O(M^3)$.

System (\ref{eq:sub_sys}) is made of $M$ first order linear differential equations, and a good
numerical solution (for instance by a Runge-Kutta method) at the cost $O(M^2)$ ($O(M)$ if $U^{(k)}$ is a sparse matrix) could be looked for. Nevertheless, this is not completely true since a numerical approach requires the use of $A(0)$, and each component of  the latter is a matrix  of order $2^M$; therefore, the linear system (\ref{eq:sub_sys}) actually involves $M2^{2M}$ differential equations.

We now discuss a different point of view for the same problem that bypasses both the need of directly computing the exponential of the matrix $U^{(k)} t$, and the huge amount of computation to obtain a numerical solution of (\ref{eq:sub_sys}), according to the classical representation of $M$-mode fermionic operators. 

Let us reconsider the system (\ref{eq:sub_sys}) but assuming that  each component $A_j(t)$ of $A$ is an $M$-component row vector whose value at $t=0$ is
the $j$-th element of the canonical orthonormal basis of $\mathbb{R}^M$.
As a consequence,  (\ref{eq:sub_sys}) now represents a linear differential equation for the $M\times M$ matrix $A$ to be solved with the initial condition $A(0)=\Id$, and now the solution reads
\begin{equation}
\label{solX}
A(t)=\exp(U^{(k)}t);
\end{equation}
we can compute this solution by numerically solving a system of $M^2$ linear differential equations, with computational cost $O(M^3)$ that reduces to $O(M^2)$ if the matrix $U^{(k)}$ is sparse (this usually occurs in
operatorial models on a lattice \cite{BO_migration,BO_ecomod,BCO_desert}).

The expression $\left|A_{jk}(t)\right|^2$ provides the mean value $n_j(t)$ corresponding to the initial values $n_\ell(0)=\delta_{k\ell}$ $(\ell=1,\ldots,M)$.
Therefore, in correspondence to the general initial values ($n_1,\ldots,n_M)$, the mean values are obtained
by means of the formula
\begin{equation}
\label{nX}
n_\ell(t)=\left|A_{\ell,1}(t)\right|^2n_1+\left|A_{\ell,2}(t)\right|^2n_2+\cdots\left|A_{\ell,N}(t)\right|^2n_M.
\end{equation}

We point out that this approach in the case where $U^{(k)}$ is not a sparse matrix does not  lead
to a relevant lowering of computation complexity. Nevertheless, this strategy becomes essential if the evolution of the system is governed by a Hamiltonian operator together with the periodic application of a rule modifying the values of some of the parameters involved in the model on the basis of the current state of the system. In fact, 
with such an approach, we have
\begin{equation}
A(t)=\exp(U^{(k)}t),\qquad t \in [(k-1)\tau,k\tau].
\end{equation}
Then, glueing the solutions in all the subintervals, we find
\begin{equation}
\label{eq:stepwise}
A(t)=\left\{
\begin{array}{lll}
\exp(U^{(1)}t) & & t \in [0,\tau],\\
\exp(U^{(2)}(t-\tau))\exp(U^{(1)}\tau) & & t \in [\tau,2\tau],\\
\exp(U^{(3)}(t-2\tau))\exp(U^{(2)}\tau)\exp(U^{(1)}\tau) & & t \in [2\tau,3\tau],\\
\ldots & & \ldots
\end{array}
\right.
\end{equation}

This kind of rule-induced stepwise dynamics clearly may generate discontinuities in the first order derivatives of the operators, but prevents the occurrence of jumps in their evolutions and, consequently, in the mean values of the number operators.
By adopting this rule, we are implicitly considering the possibility of having a time dependent Hamiltonian. However, the time dependence is, in our case, of a very special form: in each interval $[(k-1)\tau,k\tau[$ the Hamiltonian does not depend on time, but in $k\tau$ some changes may occur, according to how the system is evolving. For this reason, our Hamiltonian can be considered \emph{piecewise constant in time}. A comparison of this approach with that related to an explicitly time dependent Hamiltonian is discussed in Appendix 1.

\section{A two-mode system}
\label{sect3}

In this Section, by considering a very simple toy model introduced in \cite{BO_migration}, we will discuss the possibility of getting, in a simple and natural way, a dynamics approaching an asymptotic equilibrium state.

Let us consider a system $\mathcal{S}$, having two (fermionic) degrees of freedom, and the Hamiltonian
\begin{equation}
H=H_0+\lambda H_I,\qquad H_0=\omega_1a_1^\dagger a_1+\omega_2a_2^\dagger a_2, \quad H_I=a_1^\dagger a_2+a_2^\dagger a_1,
\label{MM23}
\end{equation}
where $\omega_j$ and $\lambda$ are real (and positive) quantities in order to ensure that $H$ is self-adjoint. The operators $a_j$ and $a_j^\dagger$ are assumed to satisfy the following anticommutation rules:
\begin{equation}
\{a_i,a_j^\dagger\}=\delta_{i,j}\,\Id,\qquad \{a_i,a_j\}=\{a_i^\dagger,a_j^\dagger\}=0,
\label{MM21}
\end{equation}
$i,j=1,2$, where, as usual, $\Id$ is the identity operator and $\{x,y\}:=xy+yx$.
Of course, when $\lambda=0$, there is no contribution in $H$ due to mutual interaction.

The eigenstates of the number operators $\hat n_j:=a_j^\dagger a_j$ are easily obtained: if $\varphi_{0,0}$ is the \emph{ground vector} of $\mathcal{S}$,
$a_1\varphi_{0,0}=a_2\varphi_{0,0}=0$, an orthonormal basis of our four-dimensional Hilbert space $\mathcal{H}$ is given by
\begin{equation}
\varphi_{0,0},\qquad \varphi_{1,0}:=a_1^\dagger\varphi_{0,0}, \qquad \varphi_{0,1}:=a_2^\dagger\varphi_{0,0}, \qquad \varphi_{1,1}:=a_1^\dagger a_2^\dagger\varphi_{0,0}.
\end{equation}
We have 
\begin{equation} 
\hat n_1\varphi_{n_1,n_2}=n_1\varphi_{n_1,n_2},\qquad \hat
n_2\varphi_{n_1,n_2}=n_2\varphi_{n_1,n_2}. \label{MM22} 
\end{equation}

The equations of motion for the annihilation operators $a_j(t)$ are
\begin{equation}
\dot a_1(t)=-i\omega_1 a_1(t)-i\lambda a_2(t),\qquad
\dot a_2(t)=-i\omega_2 a_2(t)-i\lambda a_1(t),
\label{MM24}
\end{equation}
that can be solved with the initial conditions $a_1(0)=a_1$ and $a_2(0)=a_2$. The solution looks like
\begin{equation}
\begin{aligned}
&a_1(t)=\frac{1}{2\delta}\left(a_1\left((\omega_1-\omega_2)\Phi_-(t)+\delta\Phi_+(t)\right)+2\lambda a_2\Phi_-(t)\right),\\
&a_2(t)=\frac{1}{2\delta}\left(a_2\left(-(\omega_1-\omega_2)\Phi_-(t)+\delta\Phi_+(t)\right)+2\lambda a_1\Phi_-(t)\right),
\end{aligned}
\label{MM25}
\end{equation}
where
\begin{equation}
\begin{aligned}
&\delta=\sqrt{(\omega_1-\omega_2)^2+4\lambda^2},\\
&\Phi_+(t)=2\exp\left(-\frac{it(\omega_1+\omega_2)}{2}\right)\cos\left(\frac{\delta t}{2}\right),\\
&\Phi_-(t)=-2i\exp\left(-\frac{it(\omega_1+\omega_2)}{2}\right)\sin\left(\frac{\delta t}{2}\right).
\end{aligned}
\end{equation}
Then, the functions $n_j(t):=\left<\varphi_{n_1,n_2},
\hat n_j(t)\varphi_{n_1,n_2}\right>$ are
\begin{equation}
\label{MM26-27}
\begin{aligned}
&n_1(t)=\frac{n_1(\omega_1-\omega_2)^2}{\delta^2}+ \frac{4\lambda^2}{\delta^2}
\left(n_1\cos^2\left(\frac{\delta t}{2}\right)+n_2\sin^2\left(\frac{\delta t}{2}\right)\right), \\
&n_2(t)=\frac{n_2(\omega_1-\omega_2)^2}{\delta^2}+ \frac{4\lambda^2}{\delta^2}
\left(n_2\cos^2\left(\frac{\delta t}{2}\right)+n_1\sin^2\left(\frac{\delta t}{2}\right)\right).
\end{aligned}
\end{equation}
These functions can be interpreted, in agreement with other applications considered along the years, as the densities of two species, $\mathcal{S}_1$ and $\mathcal{S}_2$, interacting as in (\ref{MM23}) in a given (small) region. The interaction Hamiltonian $H_I$ in (\ref{MM23}) describes a sort of predator-prey mechanism, and this reflects in the solution (\ref{MM26-27}), showing how the two densities, because of the interaction between  $\mathcal{S}_1$ and $\mathcal{S}_2$, oscillate in the interval $[0,1]$. Otherwise, if $\lambda=0$, $n_j(t)=n_j$, the densities stay constant, and nothing interesting happens in $\mathcal{S}$. We observe that the formulae in (\ref{MM26-27})
automatically imply that $n_1(t)+n_2(t)=n_1+n_2$, independently of $t$ and $\lambda$: the oscillations are such that they sum up to zero. We refer to \cite{BO_migration} for more details on this model, and for its role in modeling  migration, which is achieved considering a 2D version of the $H$ in (\ref{MM23}), with an additional term accounting for the diffusion of the two species in a lattice. Here, we exploit the possibility of getting some limiting values for $n_1(t)$ and $n_2(t)$ for large values of $t$, when $\lambda\neq0$.

The first trivial remark is that the functions $n_1(t)$ and $n_2(t)$ in (\ref{MM26-27}) do not admit any asymptotic limit, except when $n_1=n_2$ 
(or when $\lambda=0$, which is excluded here). In this case, clearly, $n_1(t)=n_2(t)=n_1=n_2$. On the other hand, if $n_1\neq n_2$, then both $n_1(t)$ and $n_2(t)$ always oscillate in time. This is not surprising since it is easy to prove that
if $\mathcal{S}$ is a system living in a finite dimensional Hilbert space, and if its dynamics is driven by a time independent, self-adjoint, quadratic Hamiltonian $\tilde H$, then its evolution is necessarily periodic or quasi-periodic. In fact,
 if $\tilde H=\tilde H^\dagger$ is an $M\times M$ matrix, then an unitary matrix $U$ exists such that $U\tilde HU^{-1}=H_d$, which is a diagonal matrix with $M$ (not necessarily different) real eigenvalues $E_1, E_2, \ldots, E_N$. Then, if $\hat n$ is an observable of $\mathcal{S}$, its time evolution can be written as
\begin{equation}
\begin{aligned}
\hat n(t)&=\exp(i\tilde Ht)\hat n \exp(-i\tilde Ht)=\\
&=U^{-1}\exp(iH_dt)\left(U\hat n U^{-1}\right) \exp(-iH_dt)U,
\end{aligned}
\end{equation}
which is periodic when all couples of eigenvalues of $\tilde H$ are commensurable, while is quasiperiodic otherwise. If $H_d$ commutes with $U\hat n U^{-1}$, then $\hat n(t)=\hat n$, and its large time behavior is clearly trivial. Otherwise, $\hat n(t)$ keeps on oscillating, and no asymptotic value is reached. This is essentially what is described by formulae (\ref{MM26-27}), where the focus is not really on the operators $\hat n_1$ and $\hat n_2$, but on their mean values.

\subsection{The rule as a map from $\mathcal{H}$ to $\mathcal{H}$ and the existence of an asymptotic value}\label{sect3a}

As we have discussed previously, what is interesting for us is to describe the existence of some asymptotic limit for our system $\mathcal{S}$. If $\mathcal{S}$ lives in a finite dimensional Hilbert space, our previous simple argument shows that no time independent, self-adjoint, Hamiltonian $\tilde H$ can make the job. Hence, we have to add some extra ingredients, or change something in the Hamiltonian. In fact, it is well known that if $\mathcal{S}$ interacts with a reservoir with an infinite number of degrees of freedom, \cite{Radmore}, such a limit can indeed be obtained. And, in fact, this possibility has been used in several applications also because it admits interesting interpretations, \cite{all1,all3}. However, from a technical point of view, this is probably not the easiest choice and, in fact, quite often one makes use of some finite dimensional effective Hamiltonian $H_{eff}$, with $H_{eff}\neq H_{eff}^\dagger$. For instance, this is what is done in optics \cite{tripf}. Here, we will show how the use of a rule, other than having a concrete meaning, can produce a realistic asymptotic value for the observables of $\mathcal{S}$.

In order to show this, we first rewrite (\ref{MM26-27}) as
\begin{equation}
N(t)=T_tN(0),
\label{NA11}
\end{equation}
where
\begin{equation*}
N(t)=\left(
\begin{array}{c}
n_1(t) \\
n_2(t) \\
\end{array}
\right), \quad
T_t=\frac{1}{\delta^2}\left(
\begin{array}{cc}
\delta^2-4\lambda^2\sin^2\left(\frac{\delta t}{2}\right) & 4\lambda^2\sin^2\left(\frac{\delta t}{2}\right) \\
4\lambda^2\sin^2\left(\frac{\delta t}{2}\right) & \delta^2-4\lambda^2\sin^2\left(\frac{\delta t}{2}\right)
\end{array}
\right).
\label{NA12}
\end{equation*}
Of course, the components of $N(t)$ return the expressions of $n_1(t)$ and $n_2(t)$ for all times.  Let us now see what happens if we insert a certain rule $\rho$ in the time evolution of the system.

Here, we can think of $\rho$ as a measure of $n_1(t)$ and $n_2(t)$ repeated at time $\tau$, $2\tau$, $3\tau$, $\ldots$ We know that performing a measure on a quantum system is a delicate operation, which modifies the system itself \cite{pascazio}. Therefore, there is no reason \emph{a priori} to say that the result of a measure at time $k\tau$ (after having measured the system at time $\tau$, $2\tau$, $\ldots, (k-1)\tau$) would be exactly the same as the one we deduce directly from (\ref{NA12}), \emph{i.e.}, $n_1(k\tau)$ and $n_2(k\tau)$, and in fact this is exactly what we are going to show now.

The first measure gives $N_1(\tau):=N(\tau)=T_\tau N(0)$.  Then, according to what is discussed in Section \ref{subsec:rhoHil}, we let the system evolve out of this new initial condition $N_1(\tau)$ for another \emph{time step}: $N_2(\tau):=T_\tau N_1(\tau)=T_\tau^2 N(0)$, and so on. It is quite natural to call the rule $\rho$ considered here a \emph{stop and go} rule: apparently, in fact, $\rho$ just stops the time evolution at $\tau$, $2\tau$, $3\tau$ and so on, and then let the time evolution start again. Of course, this can be related to some action on the state of the system. Also, it is not hard to imagine richer versions of the rule, where the new state of the system is fixed by some external action, as in Section \ref{sectADD1}.  So doing we produce a sequence
\begin{equation}
N_\ell(\tau)=T_\tau^\ell N(0),
\label{NA13}
\end{equation}
for all $\ell\geq1$.

In order to compute $N_\ell(\tau)$, and its limit for $\ell$ diverging, we first observe that $T_t$ is a self-adjoint matrix, so it can be easily diagonalized. In particular, we get
\begin{equation}
U^{-1}T_tU=\left(
\begin{array}{cc}
\lambda_1(t) & 0 \\
0 & \lambda_2(t) \\
\end{array}
\right)=:\Lambda_t,
\end{equation}
where
\begin{equation}
U=\frac{1}{\sqrt{2}}\left(
\begin{array}{cc}
1 & -1 \\
1 & 1 \\
\end{array}
\right), \;
\lambda_1(t)=1, \; \lambda_2(t)=\frac{1}{\delta^2}\left(\delta^2-8\lambda^2\sin^2\left(\frac{\delta t}{2}\right)\right).
\end{equation}
Then
\begin{equation}
T_\tau^\ell=U\Lambda_\tau^\ell U^{-1}=U\left(
\begin{array}{cc}
1 & 0 \\
0 & \lambda_2^\ell(\tau) \\
\end{array}
\right)U^{-1},
\end{equation}
so that $N_\ell(\tau)=T_\tau^\ell N(0)$ can converge if $\lambda_2^\ell(\tau)$ does converge when $\ell$ diverges.
This is what happens whenever the parameters $\delta$, $\tau$ and $\lambda$ satisfy the following inequalities:
\begin{equation}
0<8\lambda^2\sin^2\left(\frac{\delta t}{2}\right)<\delta^2.
\label{NA14}
\end{equation}
In fact, when this is true, $\lambda_2(\tau)\in]0,1[$, and, therefore, $\lim_{\ell\rightarrow\infty}\lambda_2^\ell(\tau)=0$. Hence,
\begin{equation}
\lim_{\ell\rightarrow\infty}N_l(\tau)=\left(
\begin{array}{c}
n_1(0) \\
0 \\
\end{array}
\right),
\label{NA15}
\end{equation}
which clearly shows that a non trivial equilibrium can be reached in this case. However, if the parameters do not satisfy (\ref{NA14}), the asymptotic behavior of $N_\ell(\tau)$ can be completely different. In fact, taking, for instance, $\tau=\frac{\pi}{\delta}$ and $\lambda=\sqrt{\frac{3}{8}}\,\delta$, and then fixing $\delta=1$ for simplicity, we deduce that $\lambda_2(\tau)=-2$, so that $\lim_{\ell\rightarrow\infty}|\lambda_2(\tau)|^\ell=\infty$: thus, it is evident that the role of the parameters of $H$ are in fact essential. This is also evident from inequality (\ref{NA14}), which is satisfied for all $t$ if $\lambda<(\omega_1-\omega_2)/2$. 

The conclusion of the analysis of this simple example is the following: even in the presence of a self-adjoint Hamiltonian, a simple two-modes fermionic system admits a non trivial asymptotic limit for a large range of values of the parameters of the model, at least if a \emph{stop and go} rule is assumed. However, the same rule can also produce a non converging dynamics for special choices of the parameters.

\subsection{The rule working on the space of parameters}
As we have discussed in Section \ref{sect2}, using a rule which modifies the state of the system is not the unique choice. We can also modify the parameters of the Hamiltonian, which is the choice we are going to discuss here. Then,
let us consider again the model with two fermionic modes ruled by the Hamiltonian (\ref{MM23}). In this case, let us
consider the $(H,\tilde\rho)$--induced dynamics where the rule $\tilde\rho$ acts on the space of parameters. More precisely, let us
assume that the rule modifies at fixed times $k\tau$  only the parameters $\omega_1$ and $\omega_2$ in (\ref{MM23}) according to the variations of $n_1$ and $n_2$ in the interval $[0,k\tau]$ (rule $\tilde\rho_A$) or in the (smaller) interval $[(k-1)\tau,k\tau]$ (rule $\tilde\rho_B$). Both these rules, see (\ref{ruleA}) and (\ref{ruleB}), are clearly completely different from a stop and go rule.

\begin{figure}
\begin{center}
\includegraphics[width=0.6\textwidth]{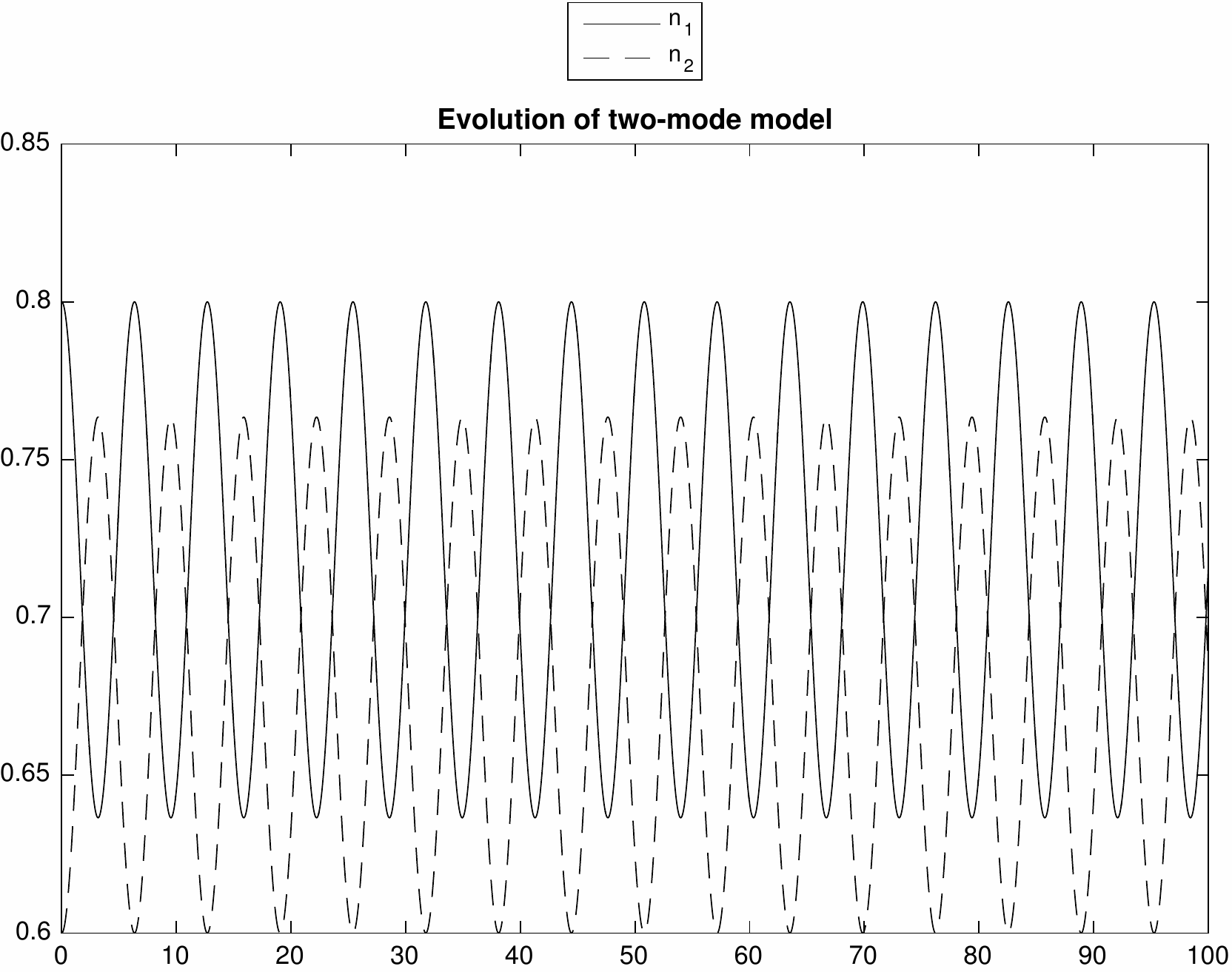}
\end{center}
\caption{\label{fig:2mode} Time evolution of the mean values: oscillating behavior.}
\end{figure}

Let us fix the initial values of the parameters, say  $\omega_1=1/\sqrt{3}$, $\omega_2=1$, $\lambda=1/\sqrt{5}$, and the
initial conditions $n_1=0.8$, $n_2=0.6$. Without applying any rule we have (see Figure~\ref{fig:2mode}), as one expects, a never ending oscillating behavior of both $n_1(t)$ and $n_2(t)$.

\begin{figure}
\begin{center}
\subfigure[$\tilde\rho_A$, $\tau=1$]{\includegraphics[width=0.48\textwidth]{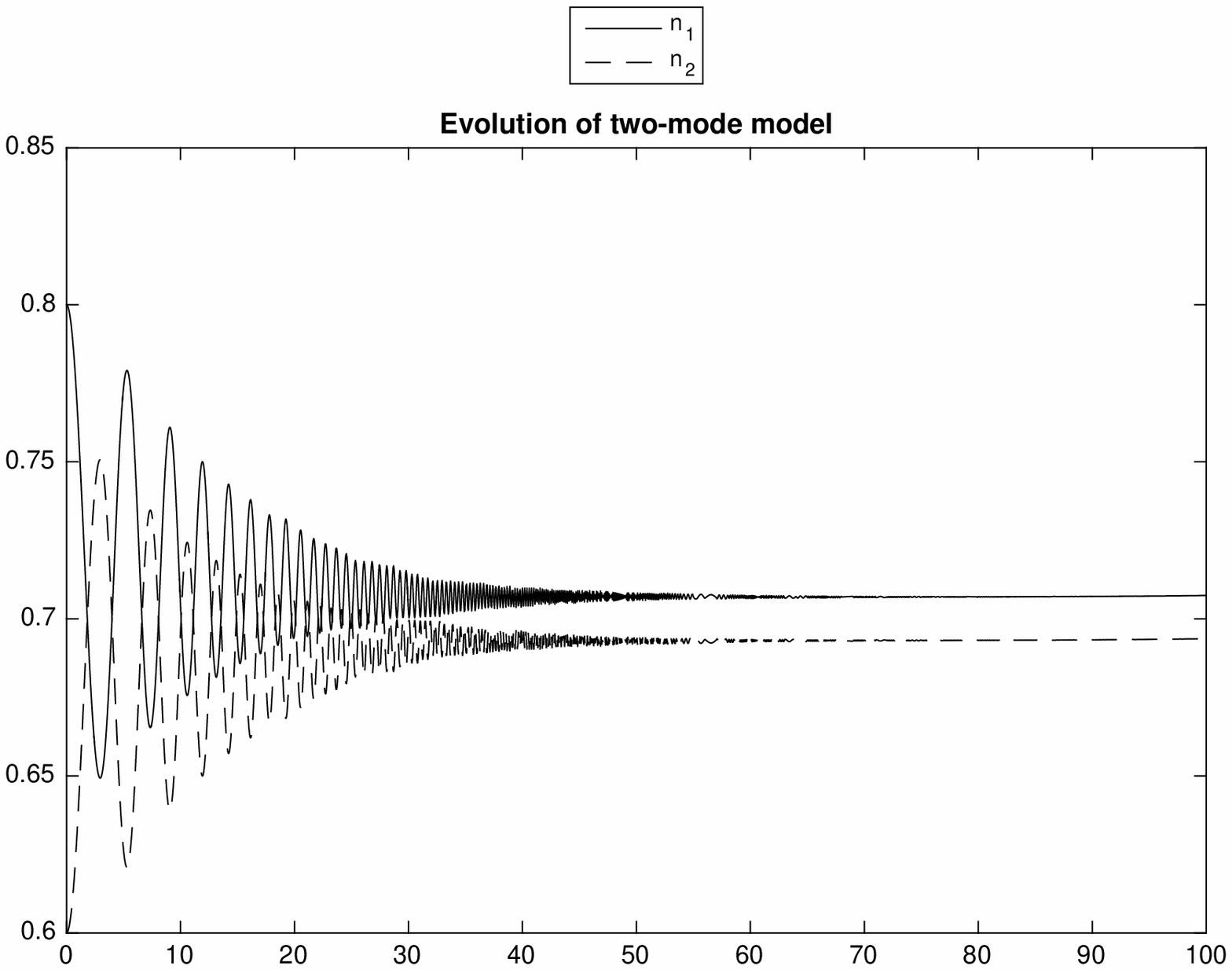}}
\subfigure[$\tilde\rho_B$, $\tau=1$]{\includegraphics[width=0.48\textwidth]{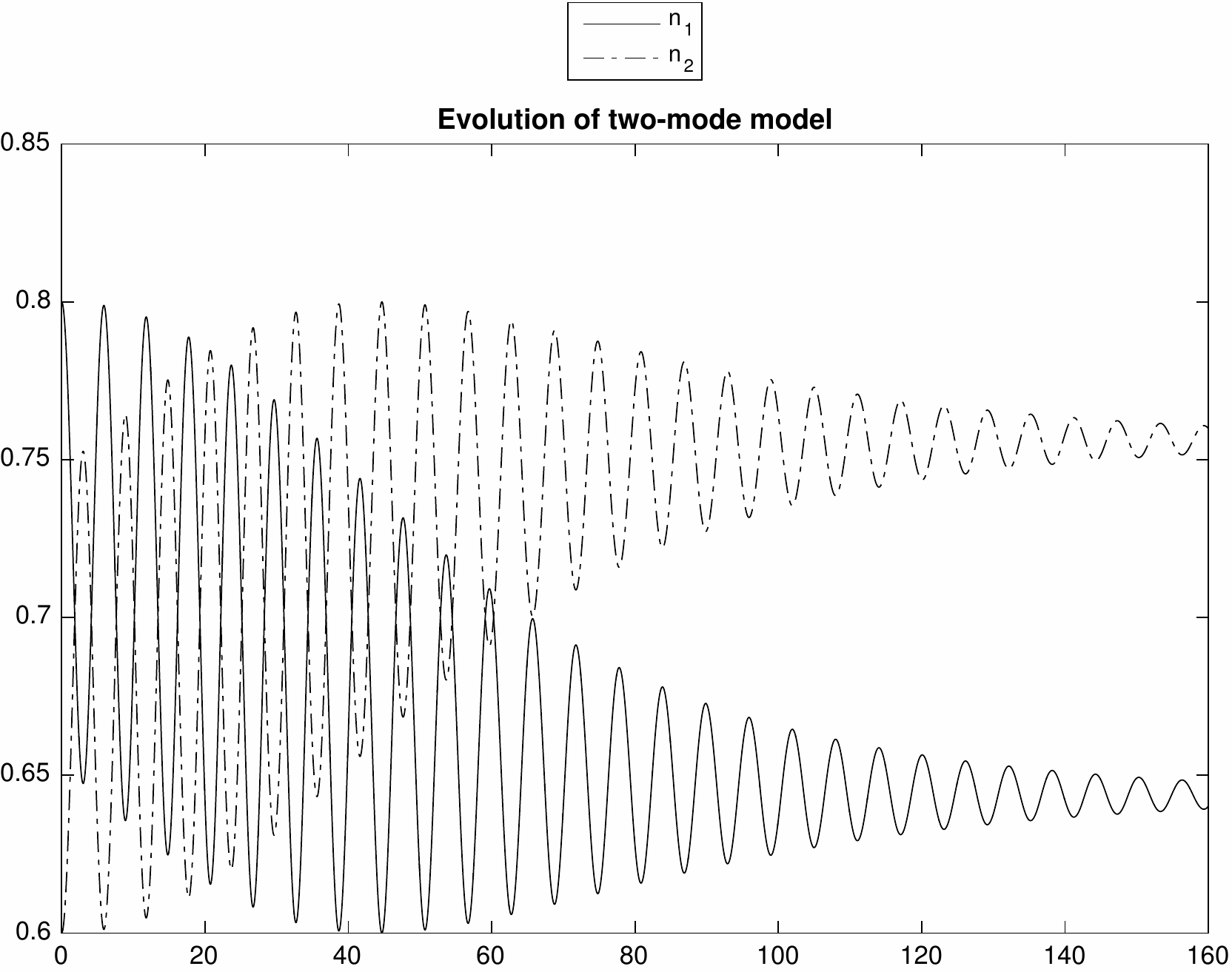}}\\
\subfigure[$\tilde\rho_A$, $\tau=2$]{\includegraphics[width=0.48\textwidth]{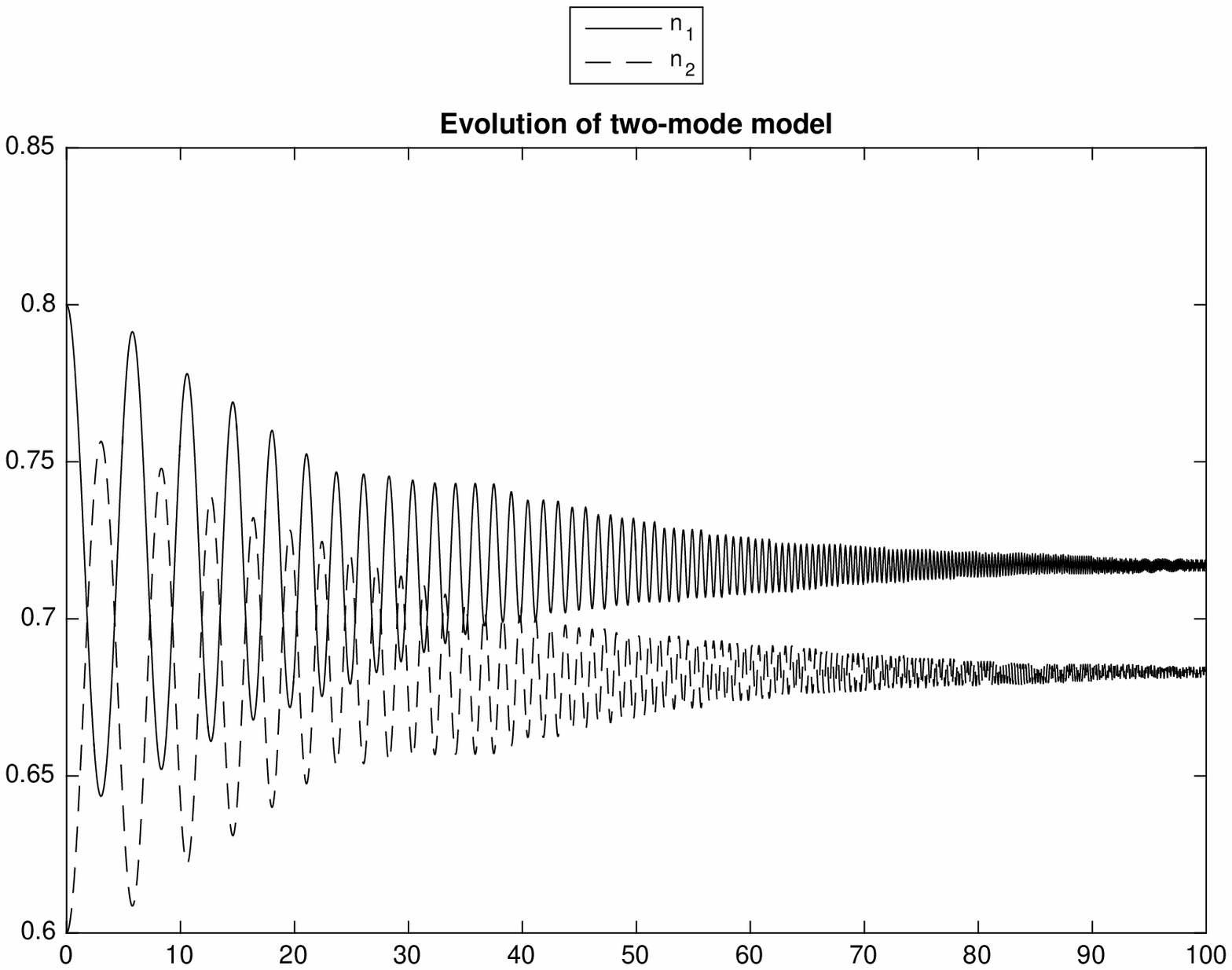}}
\subfigure[$\tilde\rho_B$, $\tau=2$]{\includegraphics[width=0.48\textwidth]{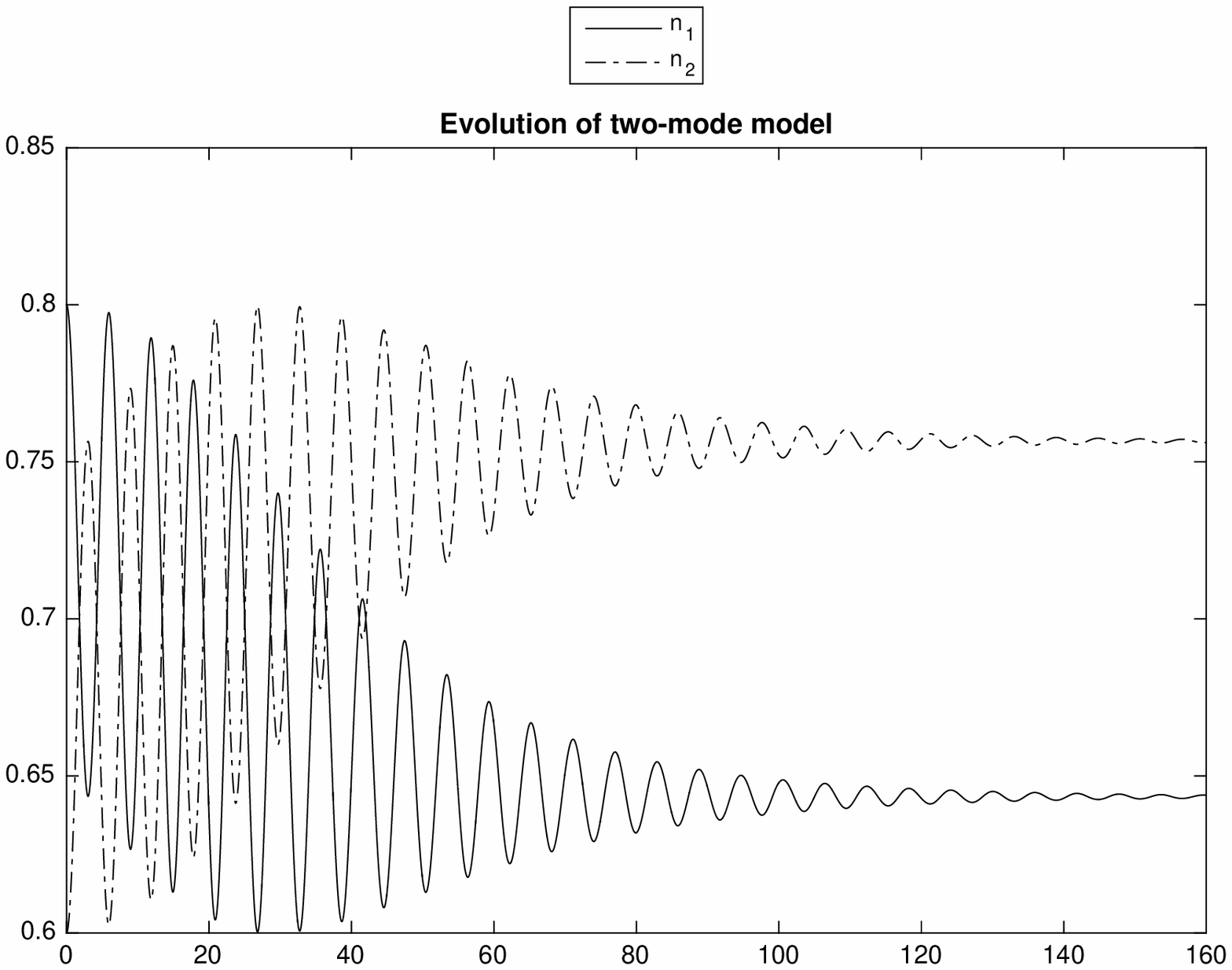}}\\
\subfigure[$\tilde\rho_A$, $\tau=4$]{\includegraphics[width=0.48\textwidth]{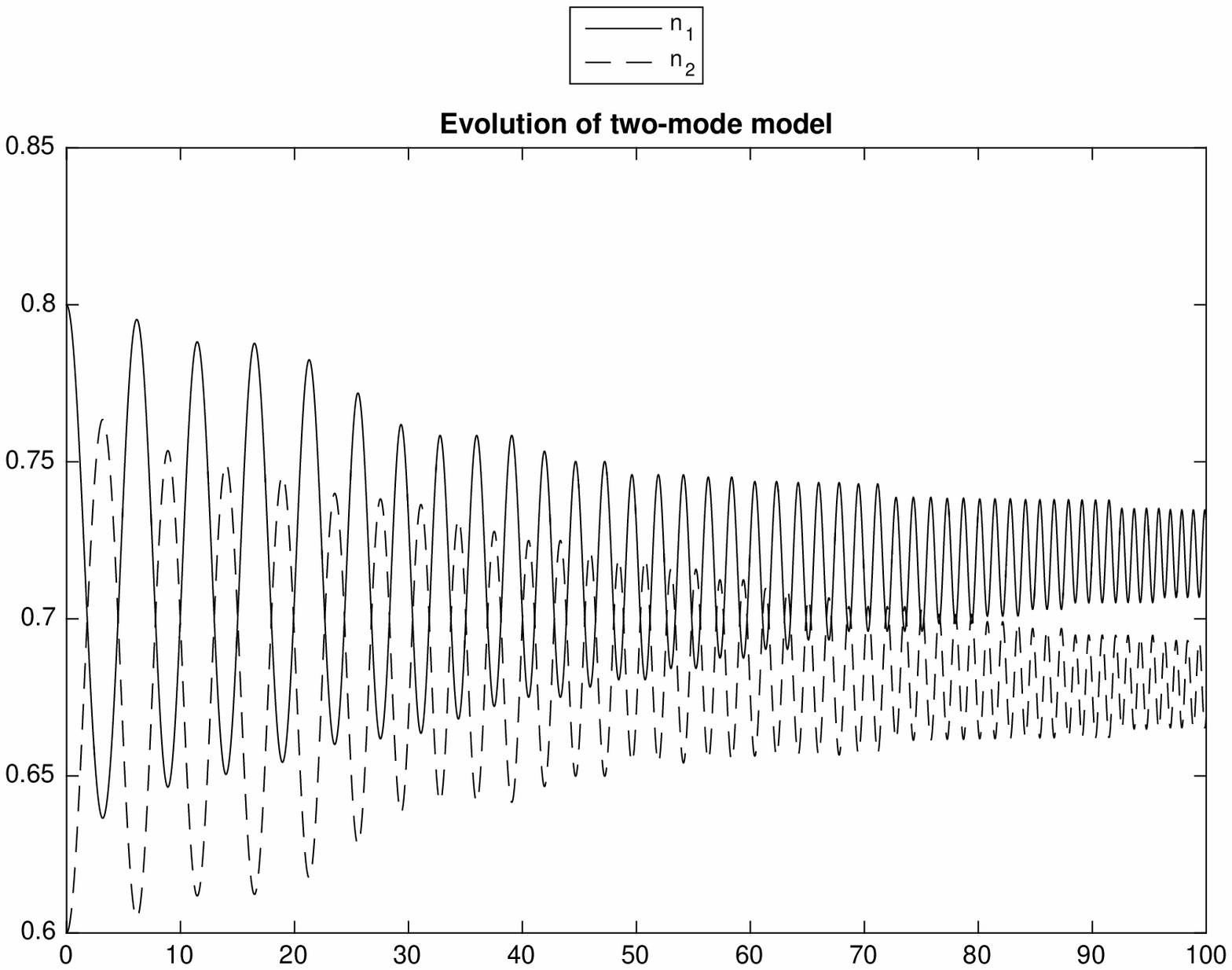}}
\subfigure[$\tilde\rho_B$, $\tau=4$]{\includegraphics[width=0.48\textwidth]{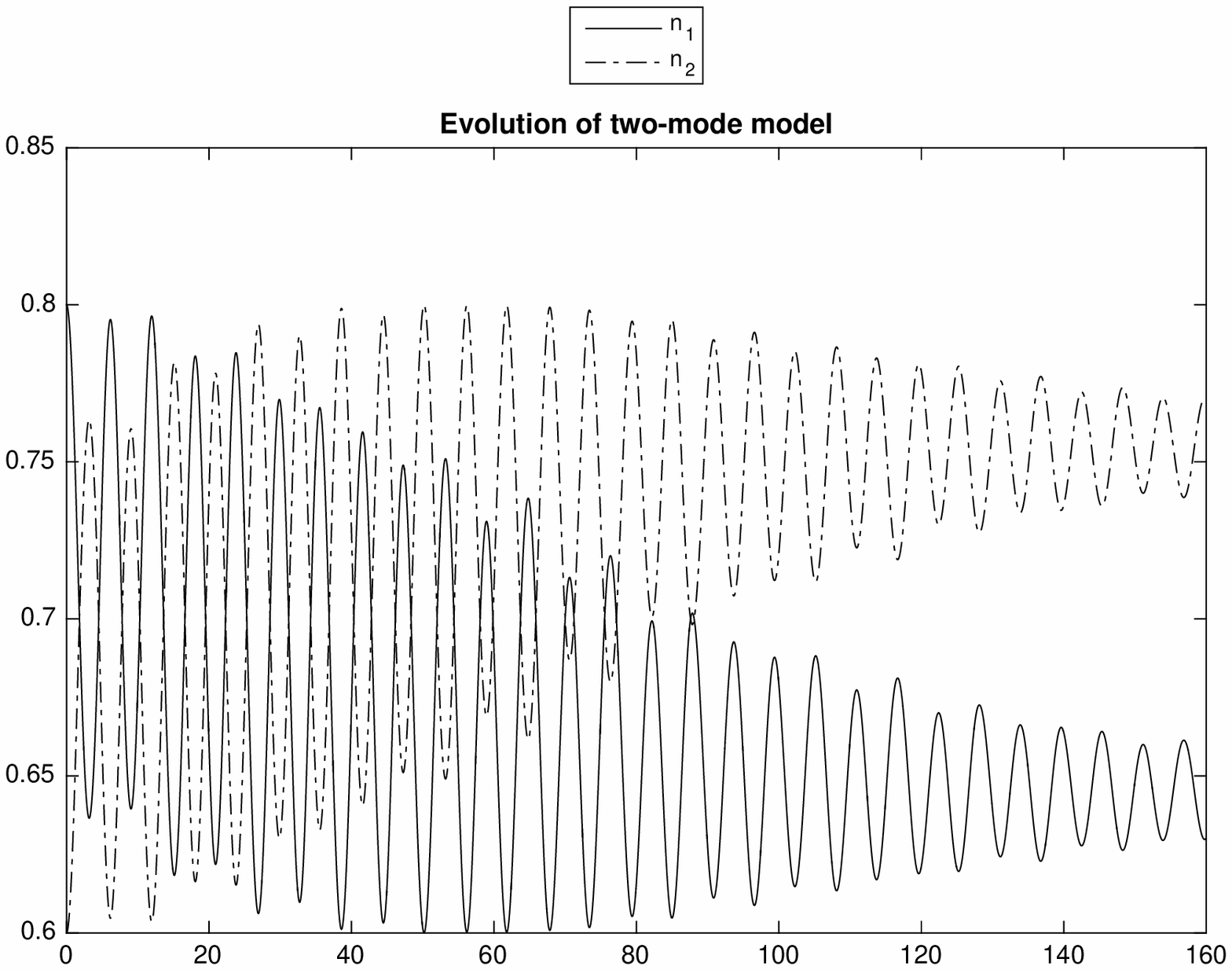}}
\end{center}
\caption{\label{fig:2mode_ruleAB}Time evolution with rule $\tilde\rho_A$ (on the left) and 
rule $\tilde\rho_B$ (on the right) for different choices of $\tau$.}
\end{figure}

By taking the rule
\begin{equation}
\label{ruleA}
\begin{aligned}
&\tilde\rho_A(\omega_1)=\omega_1(1+\delta_1), \qquad &&\delta_1=n_1(k\tau)-n_1(0),\\
&\tilde\rho_A(\omega_2)=\omega_2(1+\delta_2), \qquad &&\delta_2=n_2(k\tau)-n_2(0),
\end{aligned}
\end{equation}
we can see in the left part of Fig.~\ref{fig:2mode_ruleAB} how the system reaches some asymptotic states; also, we see that the rate of decay
of oscillating behaviors is smaller as the value of $\tau$ is increased. Notice that, since $n_1(t)+n_2(t)$ is a constant,
$\delta_1+\delta_2=0$, so that the variations of the inertia parameters $\omega_1$ and $\omega_2$ are opposite.

If we consider the rule
\begin{equation}
\label{ruleB}
\begin{aligned}
&\tilde\rho_B(\omega_1)=\omega_1(1+\delta_1), \qquad &&\delta_1=n_1(k\tau)-n_1((k-1)\tau),\\
&\tilde\rho_B(\omega_2)=\omega_2(1+\delta_2), \qquad &&\delta_2=n_2(k\tau)-n_2((k-1)\tau),
\end{aligned}
\end{equation}
we can see in the right part of Fig.~\ref{fig:2mode_ruleAB} how the system tends to reach again some asymptotic states, but in a slightly different way.

The conclusion of the analysis of this simple model is that the rules, both the one defined on $\mathcal{H}$ and those working on the space of the parameters, strongly affect the time evolution of the system producing serious consequences. For this reason, we believe they can be a valid alternative to the open system procedure  adopted, for instance, in \cite{bagbook,all1}.

\section{A many-mode system: the long term survival of bacterial populations in a square region}
\label{sect4}

In this Section, we consider an operatorial model, which takes some ideas from the one developed in \cite{BO_ecomod}, and  investigated in \cite{DSO_RM2016,DSO_AAPP2016},
for the description in a finite 2D region of  bacterial populations belonging to certain genera, such as \emph{Bacillus},  \emph{Clostridium} or \emph{Pseudomonas}, whose metabolism ensures their long term survival  in terms of latent life when under negative stimuli. In recent years, special attention has been devoted by microbiologists to \emph{Pseudomonas aeruginosa}, an ubiquitous bacterium capable to use more than one hundred of chemical compounds as carbon and energy sources, and, due to its wide metabolic versatility, to persist for prolonged periods of time without external sources of nutrients
\cite{Givskov,Overbeek,Cabral,Eberl,Carnazza2007}. \emph{Pseudomonas aeruginosa} is the ethiological agent of several diseases, and represents an emergence in hospital-acquired infections, because of its high degree of resistance against several classes of antibiotics and of its persistence also in disinfecting solutions.

\begin{figure}
\begin{center}
\includegraphics[width=0.4\textwidth]{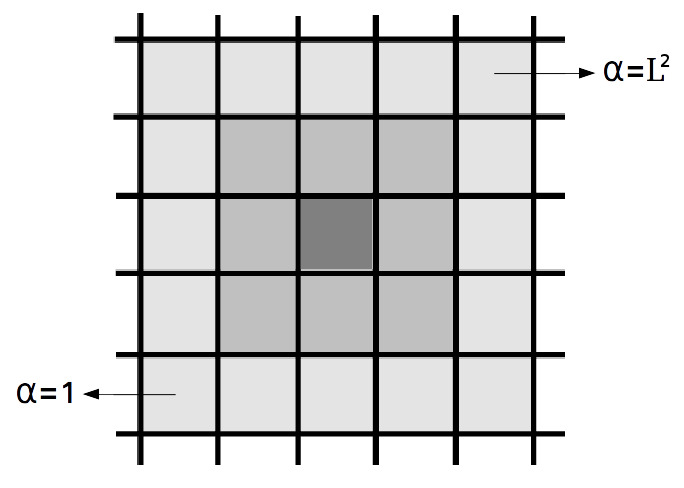}\quad
\includegraphics[width=0.4\textwidth]{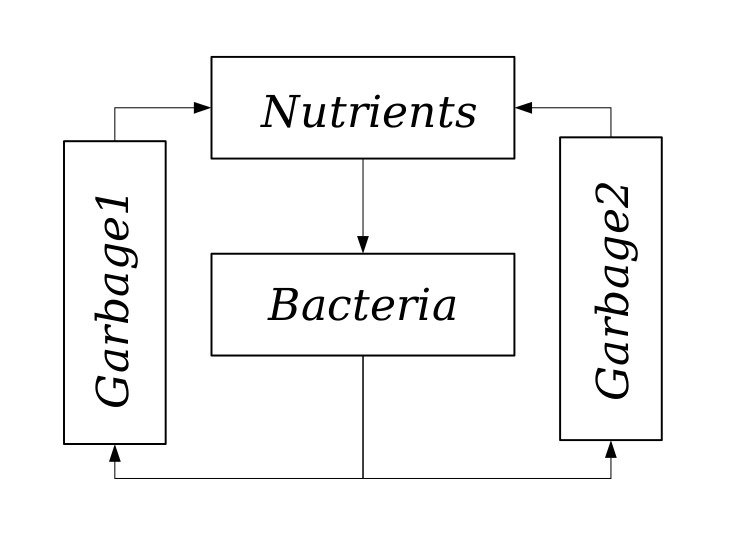}
\caption{\label{fig:lattice-model} A two-dimensional square lattice of size $L^2$ (on the left); schematic view to the $4$-compartment model of a closed ecosystem in each cell (on the right).}
\end{center}
\end{figure}

To describe colony morphology in stressed/aged bacterial populations, let us consider a region represented by a regular $L\times L$ square grid (see Fig.~\ref{fig:lattice-model}). In each cell the model is made of four compartments (Fig.~\ref{fig:lattice-model}); each compartment is represented by a fermionic operator and, according to the usual interpretation, the mean values of the number operators are used to represent the local densities of the different compartments \cite{BO_ecomod}. The first mode is related to the nutrients, the second one to the bacteria, and the last two to a couple of different garbages, see below.
The reason of the introduction of two distinct compartments for the garbage is due to the different roles they play in the ecosystem; in fact, the first one is associated to the dead cells which are rapidly reusable as nutrients, whereas the second one contains the waste material which is not yet recyclable over the short term or no longer reusable at all and, therefore, may act as a a stress factor for the system \cite{BO_ecomod,DSO_AAPP2016}.

Each actor, associated to an annihilation fermionic operator $a_{j,\alpha}$ (in the following we use latin and greek indices to refer to the compartments, and to the cells, respectively) occupies the cell $\alpha$ and interacts with the other actors in the same cell and in its Moore neighborhood\footnote{The Moore neighborhood of a cell in a 2D lattice consists in the eight cells surrounding it.}. The dynamics of the system is governed by the self-adjoint quadratic Hamiltonian operator
\begin{equation}\label{eq:Ham_lattice}
\left\{
\begin{aligned}
H&=H_0+ H_I+H_M, \qquad \hbox{ with }  \\
H_0 &= \sum_{\alpha=1}^{L^2}\,\sum_{j=1}^{4}\,\omega_{j,\alpha}\, a_{j,\alpha}^\dagger\,a_{j,\alpha},\\
H_I &= \sum_{\alpha=1}^{L^2}\,\left(
\sum_{j=2}^{4}\,\lambda_{j,\alpha}(a_{1,\alpha}\,a_{j,\alpha}^\dagger+a_{j,\alpha}\,a_{1,\alpha}^\dagger)\right.\\
&+\left.\sum_{k=3}^{4}\,\nu_{k,\alpha}(a_{2,\alpha}\,a_{k,\alpha}^\dagger+a_{k,\alpha}\,a_{2,\alpha}^\dagger)
\right),\\
H_M &= \sum_{\alpha=1}^{L^2}\,\mu_{2,\alpha}\,\sum_{\beta=1}^{L^2}p_{\alpha,\beta}\,(a_{2,\alpha}\,a_{2,\beta}^\dagger+a_{2,\beta}\,a_{2,\alpha}^\dagger),
\end{aligned}
\right.
\end{equation}
in which $j=1,2,3,4$ label respectively the nutrients, the bacteria, and the garbages of the two types. We see that $H$ consists, besides the first standard part $H_0$, of  the two contributions $H_I$ and $H_M$, which model respectively the interactions among
the different actors, and the migration or diffusion of the bacteria, respectively. All the parameters involved in $H$ are real numbers having the following interpretation: the constants
$\omega_{j,\alpha}$ describe the \emph{inertia} of all the compartments in each cell (expressing the tendency of each degree of freedom to stay constant in time \cite{bagbook}), while $\lambda_{j,\alpha}$ ($j=2,3,4$), $\nu_{k,\alpha}$ ($k=3,4$) are used to characterize the strength of the interactions among the bacteria, the nutrients, and the garbages;  the remaining parameters $p_{\alpha,\beta}$ and $\mu_{2,\alpha}$ are related to the possibility of the bacterial population to move from the cell $\alpha$ to cell $\beta$, and to its mobility, respectively.

The linear equations of motion deduced for each operator in accordance with the Heisenberg scheme read
\begin{equation}
\label{eq.lattice}
\left\{
\begin{aligned}
\dot a_{1,\alpha}&=i\left(-\omega_{1,\alpha} a_{1,\alpha}+\lambda_{2,\alpha}\,a_{2,\alpha}+\lambda_{3,\alpha}\,a_{3,\alpha}+\lambda_{4,\alpha}\,a_{4,\alpha} \right),\\
\dot a_{2,\alpha}&=i \left( -\omega_{2,\alpha} a_{2,\alpha}+\lambda_{2,\alpha}\,a_{1,\alpha}+\nu_{3,\alpha}\,a_{3,\alpha}+\nu_{4,\alpha}\,a_{4,\alpha} \phantom{\sum_{\beta=1}^{L^2}}\right.\\
&+\left.\sum_{\beta=1}^{L^2}(\mu_{2,\alpha}+\mu_{2,\beta})\,p_{\alpha,\beta}\,a_{2,\beta}\right),\\
\dot a_{3,\alpha}&=i\left(-\omega_{3,\alpha}a_{3,\alpha}+\lambda_{3,\alpha}\,a_{1,\alpha}+\nu_{3,\alpha}\,a_{2,\alpha} \right),\\
\dot a_{4,\alpha}&=i\left(-\omega_{4,\alpha}a_{4,\alpha}+\lambda_{4,\alpha}\,a_{1,\alpha}+\nu_{4,\alpha}\,a_{2,\alpha} \right),
\end{aligned}
\right.
\end{equation}
$\alpha=1,\ldots,L^2$.

\begin{figure}[!]
\begin{center}
\subfigure[Bacterial densities at $t=0.5$]{\includegraphics[width=0.48\textwidth]{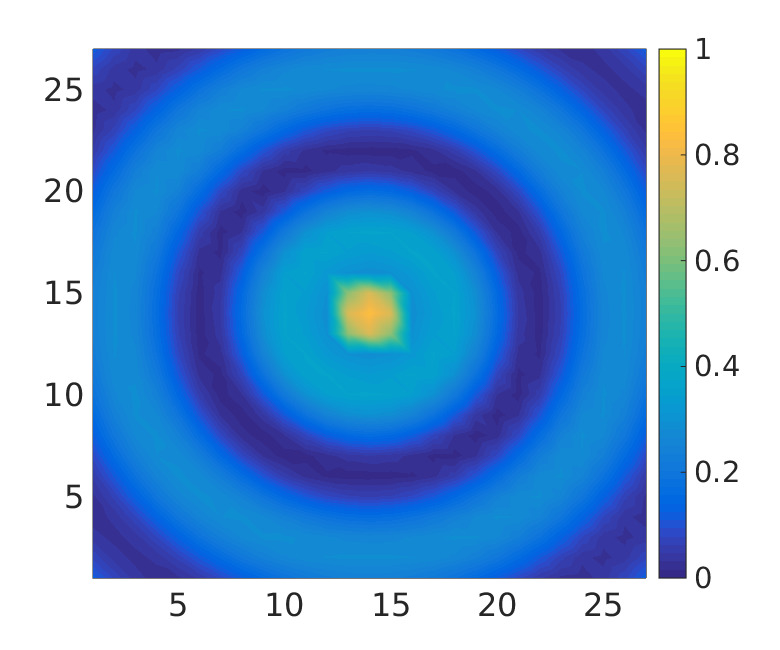}}
\subfigure[Bacterial densities at $t=1.7$]{\includegraphics[width=0.48\textwidth]{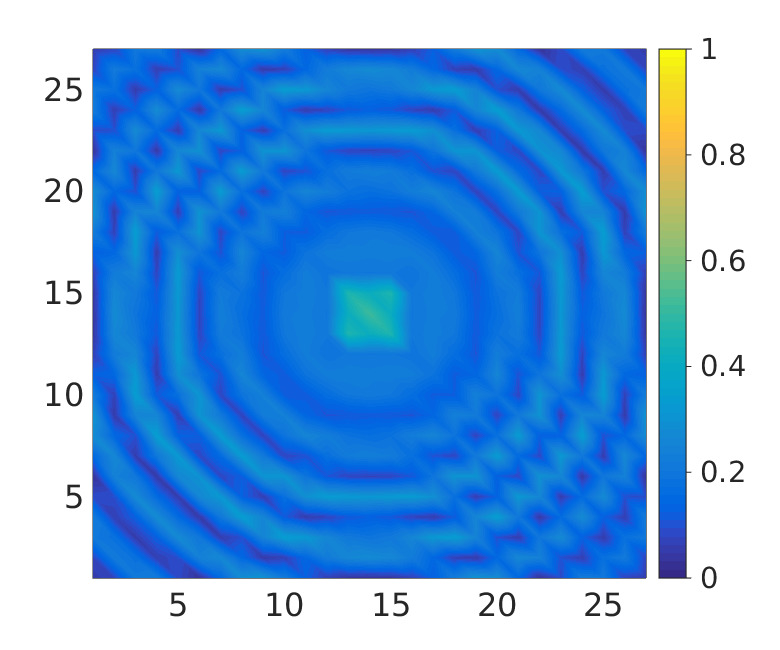}}
\subfigure[Bacterial densities at $t=3.6$]{\includegraphics[width=0.48\textwidth]{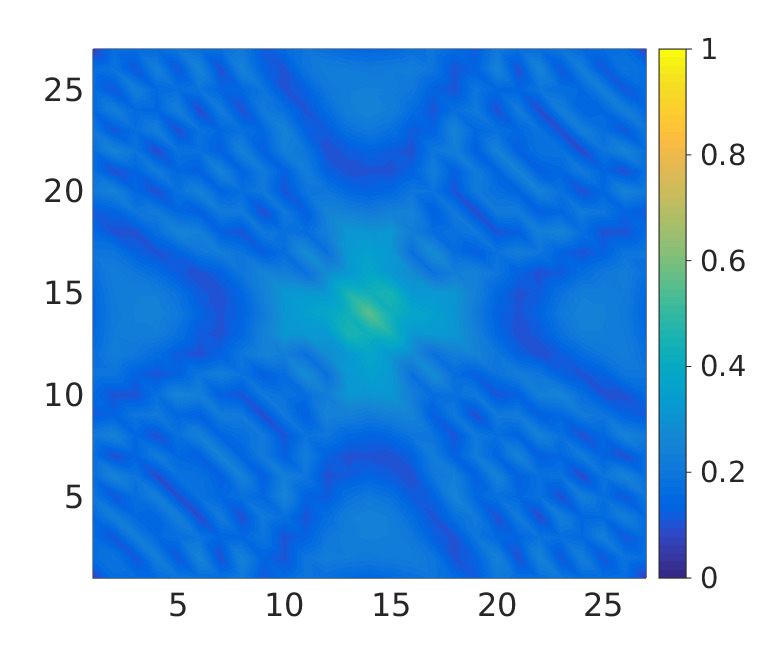}}
\subfigure[Bacterial densities at $t=5.4$]{\includegraphics[width=0.48\textwidth]{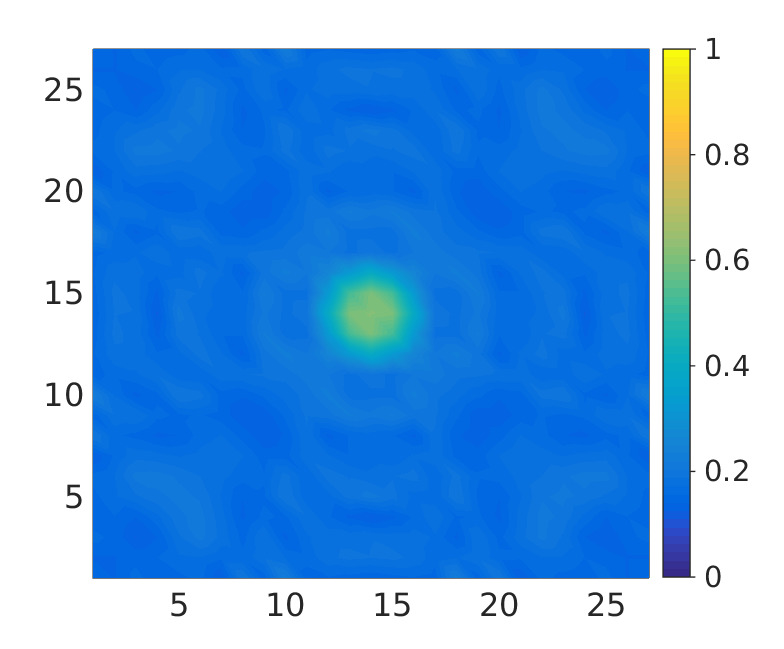}}
\end{center}
\caption{\label{fig:norules} Nonhomogeneous linear model. The frames show for each row the densities of the bacteria over the entire region at times 0.5, 1.7, 3.6, 5.4 respectively.}
\end{figure}

\begin{figure}[!]
\begin{center}
\subfigure[Mean of the densities  all over the cells]{\includegraphics[width=0.48\textwidth]{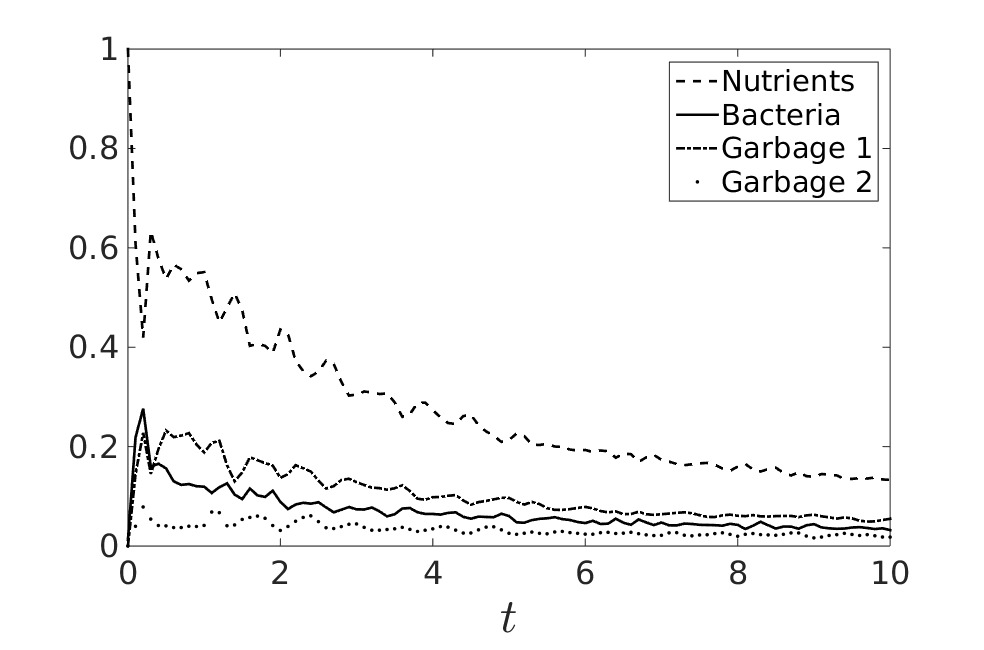}}
\subfigure[Variance of the densities  all over the cells]{\includegraphics[width=0.48\textwidth]{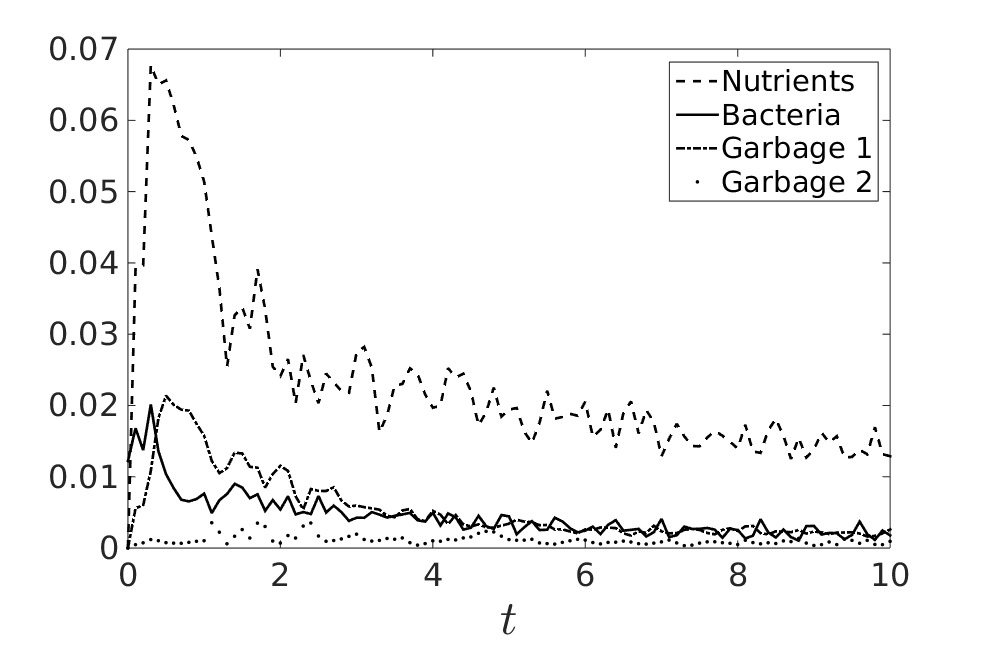}}
\end{center}
\caption{\label{fig:meanvarnorules}Time evolution of the mean (on the left)  and the variance (on the right) of the densities of all the compartments of the model over the 2D region.}
\end{figure}

Fig.~\ref{fig:norules} shows the densities of the bacteria at different instants as deduced by numerically integrating system~\eqref{eq.lattice}, and then by computing the mean values of the number operators over the initial state. We consider a nonhomogeneous region characterized by cell dependent parameters such that, when moving away from the center, the inertia parameters decrease and the interaction parameters moderately grow. Precisely, the parameters entering $H$ are chosen as in \cite{DSO_RM2016}, say, $\omega_{1,\alpha}=0.3/(1+d_\alpha)$, $\omega_{2,\alpha}=0.1/(1+d_\alpha)$, $\omega_{3,\alpha}=0.3/(1+d_\alpha)$, $\omega_{4,\alpha}=0.4/(1+d_\alpha)$,
$\lambda_{2,\alpha}=0.5 d_\alpha$, $\lambda_{3,\alpha}=0.4d_\alpha$, $\lambda_{4,\alpha}=0.2d_\alpha$, $\nu_{2,\alpha}=0.6d_\alpha$, $\nu_{3,\alpha}=0.4d_\alpha$, provided that $d_\alpha\neq 0$. Here  $d_\alpha$ is the Euclidean distance (normalized to 1) between the cell $\alpha$ and the central one; moreover,  $\mu_{2,\alpha}=0.3$ ($\alpha=1,\ldots,L^2$), whereas $p_{\alpha,\beta}=p_{\beta,\alpha}$ is vanishing for $\alpha=\beta$, equal to $1/d^2(\alpha,\beta)$, where $d(\alpha,\beta)$ is the Euclidean distance between the cell $\alpha$ and the cell $\beta$ belonging to the Moore neighborhood of cell $\alpha$, and zero elsewhere. As initial data we chose a density 1  for the nutrients  over all the cells, a density 1 for the bacteria in the 9 central cells of the lattice, and  vanishing densities for the two garbages  all over the cells. These initial conditions reflect  the fact that, at $t=0$, the nutrients are uniformly distributed, the bacteria are localized in a restricted central area, and the two garbages are both empty. In Fig.~\ref{fig:meanvarnorules}, the mean and the variance of the densities of the various compartments all over the cells are shown. It can be seen that these average densities tend
to asymptotic values. The plot of variances shows a clear effect of the diffusion, that is a progressive
homogenization of the densities.

Once we have defined a suitable functional form for the Hamiltonian operator, we now move to describe the effect of combining the usual quantum definition of the dynamics with the action of specific rules periodically acting on the system. The rules we introduce have a simple physical or biological interpretation. As in Section \ref{sect3}, we will consider two different possibilities: the first one changes the vector of the system, while the second one acts on the parameters of the Hamiltonian.

\subsection{Rules acting on the state of the system}\label{sectADD1}
As previously stated in Section \ref{subsec:rhoHil}, if
$\varphi_{\mathbf{n}^0}(0)$ is the initial state of the system,
the evolved state $\exp(-iH\tau)\varphi_{\mathbf{n}^{0}}(0)$  at time $\tau$ is mapped by the rule to a new state $\varphi_{\mathbf{n}^1}\in \mathcal{H}$, which becomes  the new initial state of the system for a new iteration, $\varphi_{\mathbf{n}^1}$.
By applying the rule several times, at equally spaced time steps, we produce a sequence of different initial states $\varphi_{\mathbf{n}^{0}}(0),\varphi_{\mathbf{n}^{1}}(0),\ldots$, which evolve always with the same Hamiltonian $H$.
This approach was originally adopted in the \textit{Quantum Game of Life} \cite{BDSGO_GoL}.

Since our model involves 4 different populations
in $L^2$ cells, we construct the elements of the basis of the $2^{4L^2}$-dimensional Hilbert space $\mathcal{H}$, spanned by
\begin{equation}
\begin{aligned}
&\varphi_{\mathbf{n_1},\mathbf{n_2},\mathbf{n_3},\mathbf{n_4}}=\\
&\quad=(a^\dagger_{1,1})^{n_{1,1}}\cdots(a^\dagger_{4,1})^{n_{4,1}}\cdots(a^\dagger_{1,L^2})^{n_{1,L^2}}\cdots(a^\dagger_{4,L^2})^{n_{4,L^2}}\varphi_{\mathbf{0_1},\mathbf{0_2},\mathbf{0_3},\mathbf{0_4}},
\end{aligned}
\end{equation}
where $\varphi_{\mathbf{0_1},\mathbf{0_2},\mathbf{0_3},\mathbf{0_4}}$ is the ground vector.
Here, $\mathbf{n_j} = (n_{j,1},\cdots,n_{j,L^2})$ is the generic $L^2$-dimensional vector relative to the $j-th$ population, and each $n_{j,\alpha}$ is 0 or 1. The initial state $\varphi_{\mathbf{n}^0}(0)$ of the system can be now expressed as
\begin{equation}
\varphi_{\mathbf{n^0}}(0)=\sum\limits_{\mathbf{n_1},\mathbf{n_2},\mathbf{n_3},\mathbf{n_4}}c^0_{\mathbf{n_1},\mathbf{n_2},\mathbf{n_3},\mathbf{n_4}}\varphi_{\mathbf{n_1},\mathbf{n_2},\mathbf{n_3},\mathbf{n_4}},
\label{initstate}
\end{equation}
where $c^0_{\mathbf{n_1},\mathbf{n_2},\mathbf{n_3},\mathbf{n_4}}$ are complex scalar. Although generally required in conservative quantum mechanics, there is no need to assume a normalized initial state, since, in general, it is not guaranteed, neither natural, that the rule we are adopting preserves the probabilistic interpretation of the states.

As an application to the bacterial population dynamics, we choose a rule such that, at the generic $k$-th iteration, the evolved state
$\varphi_{\mathbf{n}^{(k-1)}}(\tau)=\exp(-iH\tau)\varphi_{\mathbf{n}^{(k-1)}}(0)$ is mapped into a new state $\varphi_{\mathbf{n}^{(k)}}(0)$ corresponding to removing the second garbage in some specific subregion $\mathcal{O}$ of the lattice and reintroducing the same amount of garbage removed as nutrients for the bacteria, uniformly distributed all over the lattice.
More explicitly, at step $k$, we
compute the total amount of the garbage of the second type removed,
\begin{equation}
N_g=\sum\limits_{\alpha\in \mathcal{O}}n^{(k-1)}_{4,\alpha}(\tau),
\end{equation}
and we set
\begin{equation}
\label{newdens}
\begin{aligned}
&n^{(k)}_{4,\alpha}(0)=0, \quad \forall \alpha\in\mathcal{O},\\
&n^{(k)}_{1,\alpha}(0)=n^{(k-1)}_{1,\alpha}(\tau)+N_g/L^2,
\end{aligned}
\end{equation}
if $n^{(k-1)}_{1,\alpha}(\tau)+N_g/L^2\leq1$ in each cell $\alpha$.
Otherwise, we remove only that part of the garbage which ensures the condition
$n^{(k-1)}_{1,\alpha}(\tau)+N_g/L^2\leq1$ (for all $\alpha$).

The densities of bacteria and of first garbage remain unchanged in each cell,
\begin{equation}
\begin{aligned}
&n^{(k)}_{2,\alpha}(0)=n^{(k-1)}_{2,\alpha}(\tau),\quad \forall \alpha\in\mathcal{O},\\
&n^{(k)}_{2,\alpha}(0)=n^{(k-1)}_{2,\alpha}(\tau),\quad \forall \alpha\in\mathcal{O}.
\end{aligned}
\label{newdens2}
\end{equation}

It is then evident that the sum of the densities stay constant,
\begin{equation}
\sum_{i,\alpha}n^{(k)}_{i,\alpha}(0)=\sum_{i,\alpha} n^{(k-1)}_{i,\alpha}(\tau).
\end{equation}
Hence, we define the new initial state $\varphi_{\mathbf{n}^{(k)}}(0)$ fixing the coefficients in \eqref{initstate} as
\[
\begin{aligned}
&c^k_{\mathbf{\tilde n^{\alpha}_1},\mathbf{0_2},\mathbf{0_3},\mathbf{0_4}}=\sqrt{n^{k}_{1,\alpha}(0)},\qquad 
&&c^k_{\mathbf{\mathbf{0_1},\tilde n^{\alpha}_2},\mathbf{0_3},\mathbf{0_4}}=\sqrt{n^{k}_{2,\alpha}(0)},\\
&c^k_{\mathbf{\mathbf{0_1},\mathbf{0_2},\tilde n^{\alpha}_3},\mathbf{0_4}}=\sqrt{n^{k}_{3,\alpha}(0)},\qquad
&&c^k_{\mathbf{\mathbf{0_1},\mathbf{0_2},\mathbf{0_3},\tilde n^{\alpha}_4}}=\sqrt{n^{k}_{4,\alpha}(0)},
\end{aligned}
\]
where $\tilde{\mathbf{n}}^\alpha_j$, for all $j$, is the $\alpha$-th element of the canonical basis of $\mathbb{R}^{L^2}$, and $\mathbf{0}_j$ is the zero vector in $\mathbb{R}^{L^2}$. It easy to check that the mean values of the various number operators $\hat n_{j,\alpha}$ over the new initial state $\varphi_{\mathbf{n}^k}(0)$ return exactly the densities $n^{(k)}_{j,\alpha}(0)$.

Hence, the rules $\rho$ in \eqref{add2} which defines the new initial states at each new iterations are expressed through \eqref{newdens}--\eqref{newdens2}, and
\begin{equation}
\varphi_{\mathbf{n}^0}(0) \stackrel{\rho}{\longrightarrow} \varphi_{\mathbf{n}^1}(0) \stackrel{\rho}{\longrightarrow} \varphi_{\mathbf{n}^2}(0) \stackrel{\rho}{\longrightarrow} \ldots \stackrel{\rho}{\longrightarrow} \varphi_{\mathbf{n}^k}(0),
\end{equation}

In analogy with (\ref{eq:stepwise}), we can construct the  sequence of density functions
\begin{equation}
\hat n_{j,\alpha}(t)=
\left\{
\begin{array}{ll}
n^0_{j,\alpha}(t)\qquad & t\in [0,\tau[,   \\
n^1_{j,\alpha}(t-\tau)\qquad & t\in [\tau,2\tau[ , \\
n^2_{j,\alpha}(t-2\tau)\qquad & t\in [2\tau,3\tau[  , \\
\ldots &
\end{array}%
\right.
\label{DensFunc}
\end{equation}
for each $j$ and $\alpha$.
The parameters entering $H$, as well as the initial data are those used previously, in absence of the rule.

\begin{figure}[!]
\begin{center}
\subfigure[Mean of the densities all over the cells]{\includegraphics[width=0.48\textwidth]{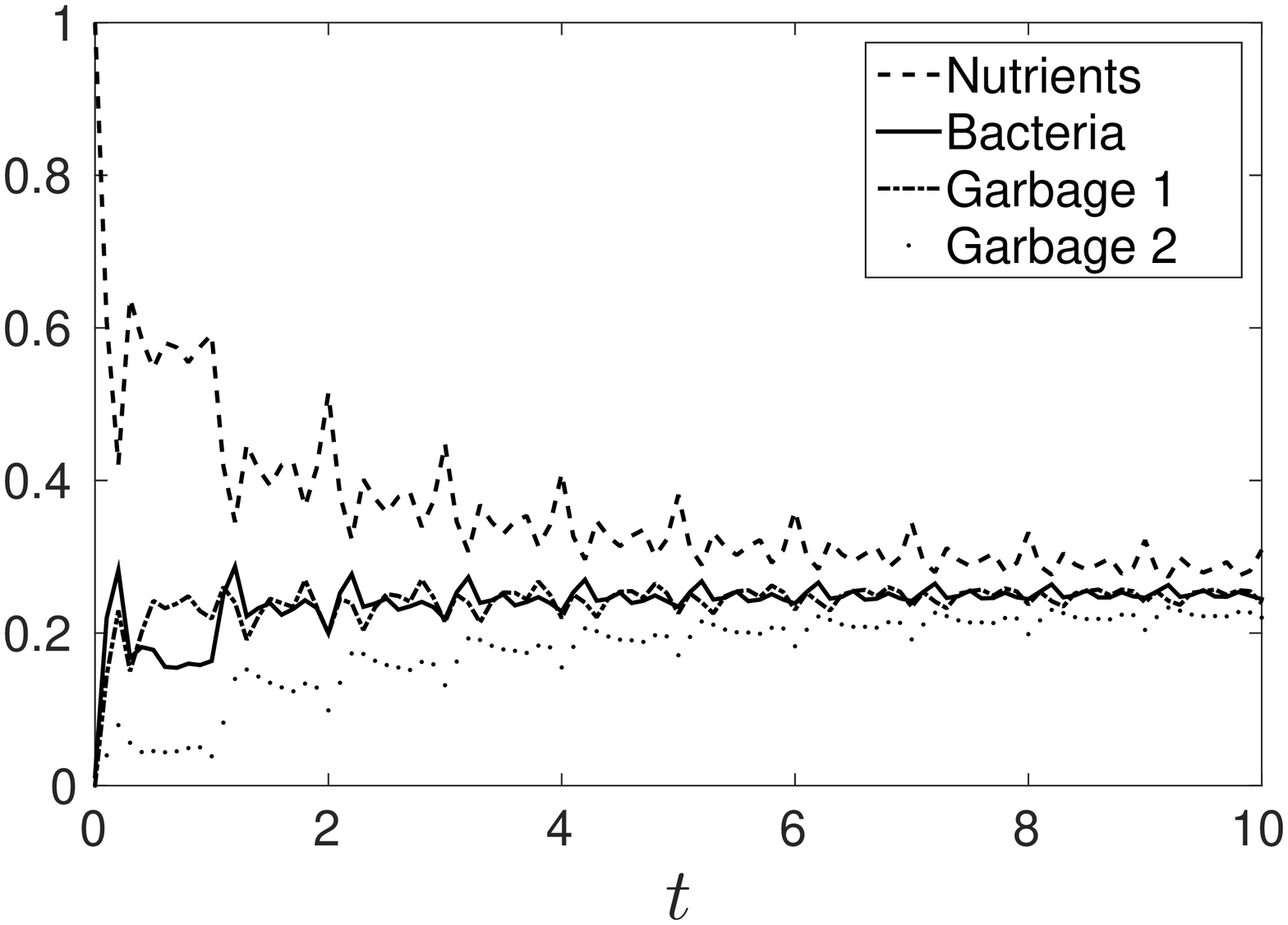}}
\subfigure[Variance of the densities all over the cells]{\includegraphics[width=0.48\textwidth]{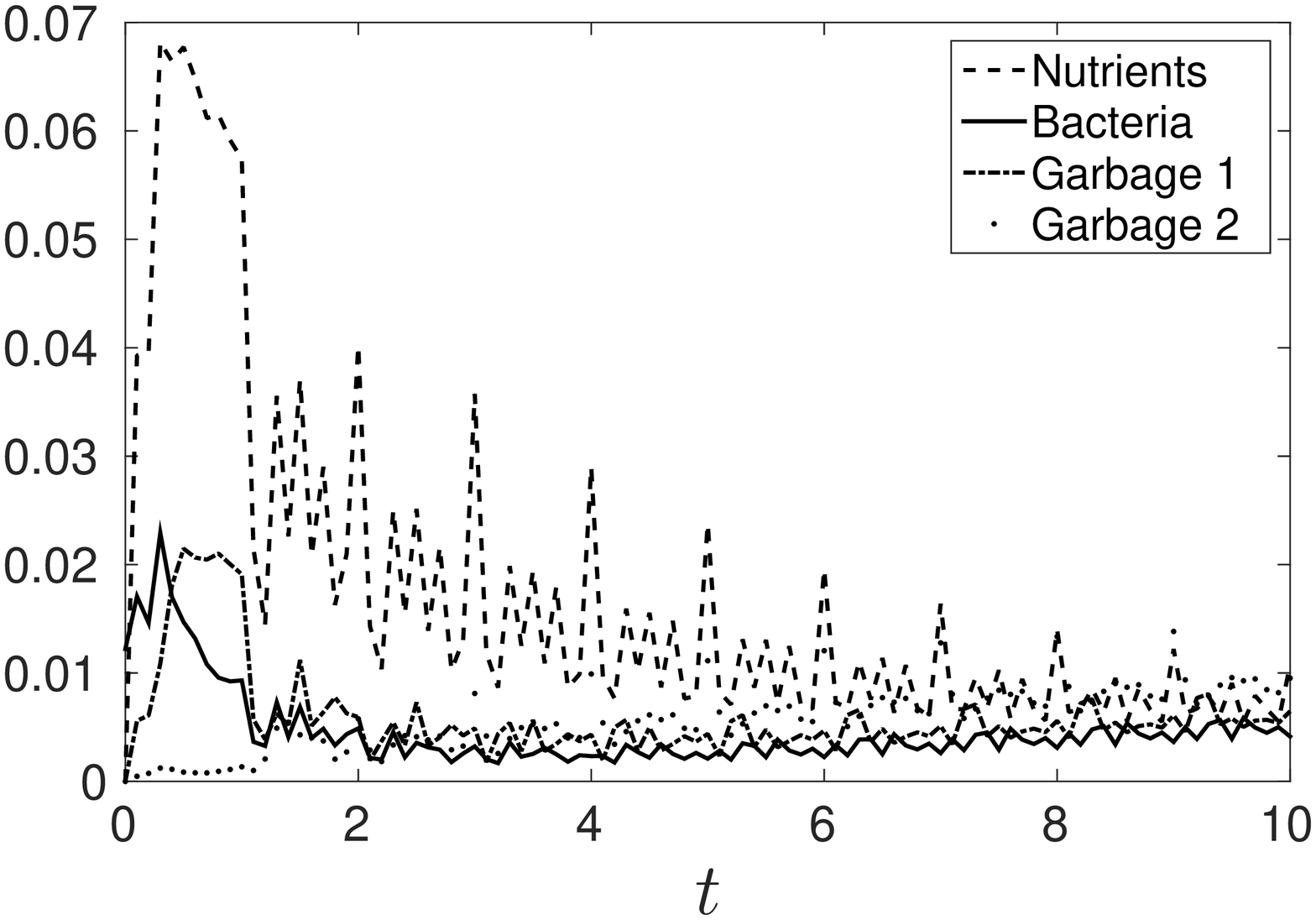}}
\end{center}
\caption{\label{fig:SG27_mean}Time evolution of the averages (on the left) and the standard deviations (on the right) of the densities of all the compartments of the ecosystem all over the cells with
$\tau=1$.}
\end{figure}

\begin{figure}[!]
\begin{center}
\vspace*{-2cm}\subfigure[Bacterial density, $t=1.7$]{\includegraphics[width=0.46\textwidth]{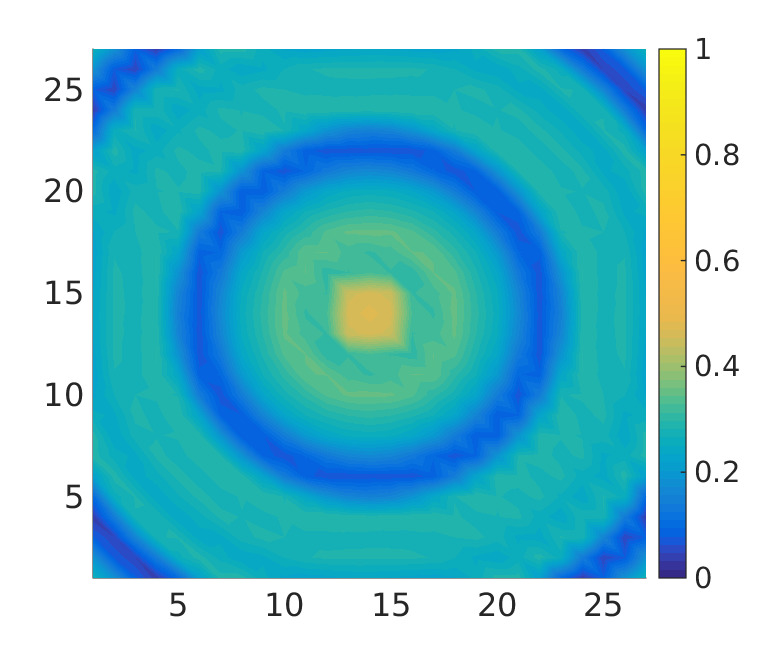}}
\subfigure[Second garbage  density, $t=1.7$]{\includegraphics[width=0.46\textwidth]{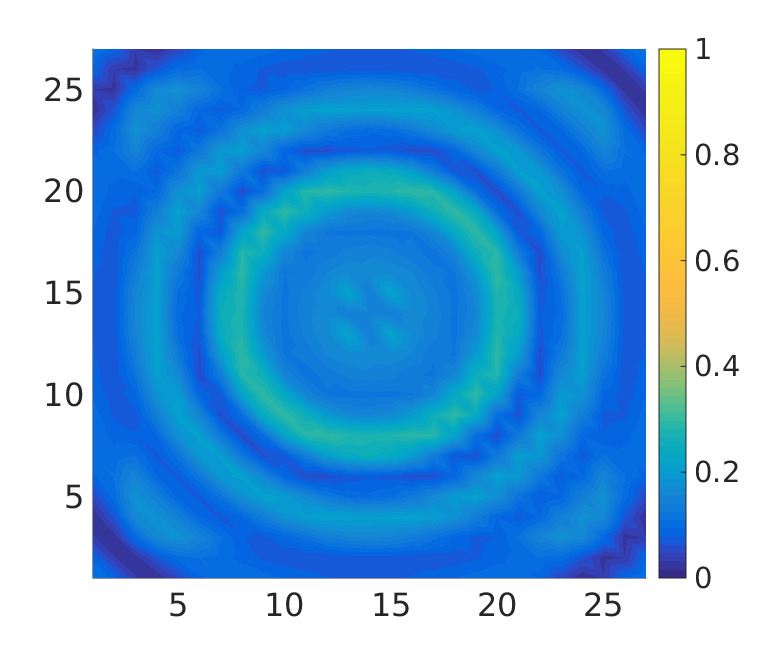}}\\
\vspace*{-0.25cm}
\subfigure[Bacterial density, $t=3.6$]{\includegraphics[width=0.46\textwidth]{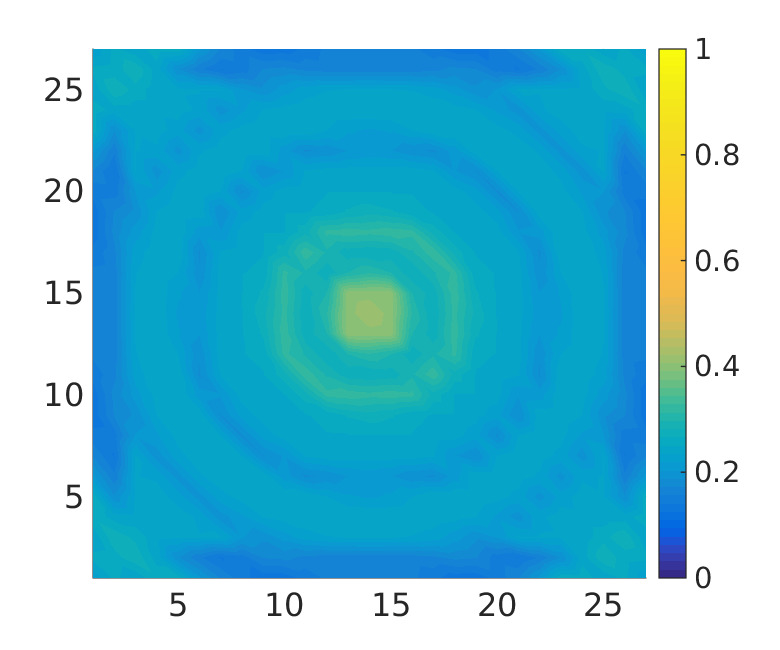}}
\subfigure[Second garbage  density, $t=3.6$]{\includegraphics[width=0.46\textwidth]{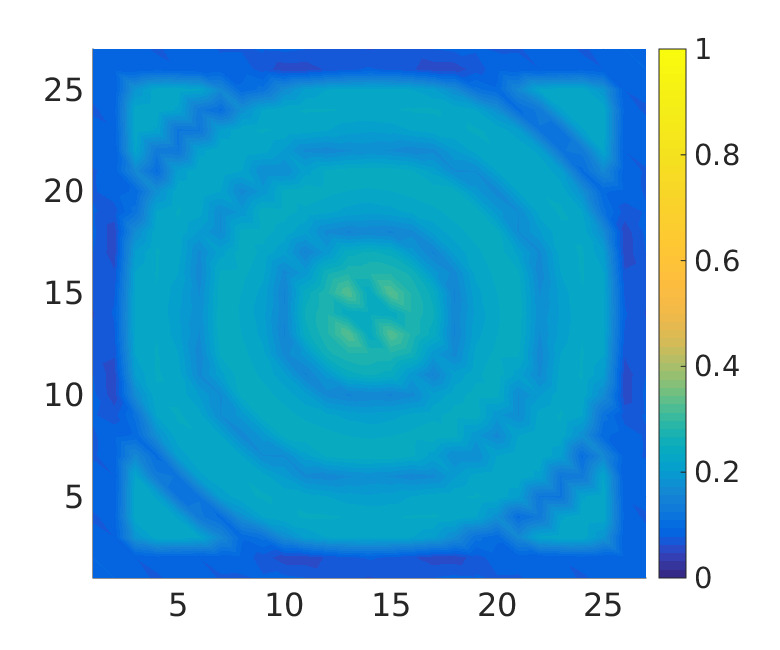}}\\
\vspace*{-0.25cm}
\subfigure[Bacterial density, $t=5.4$]{\includegraphics[width=0.46\textwidth]{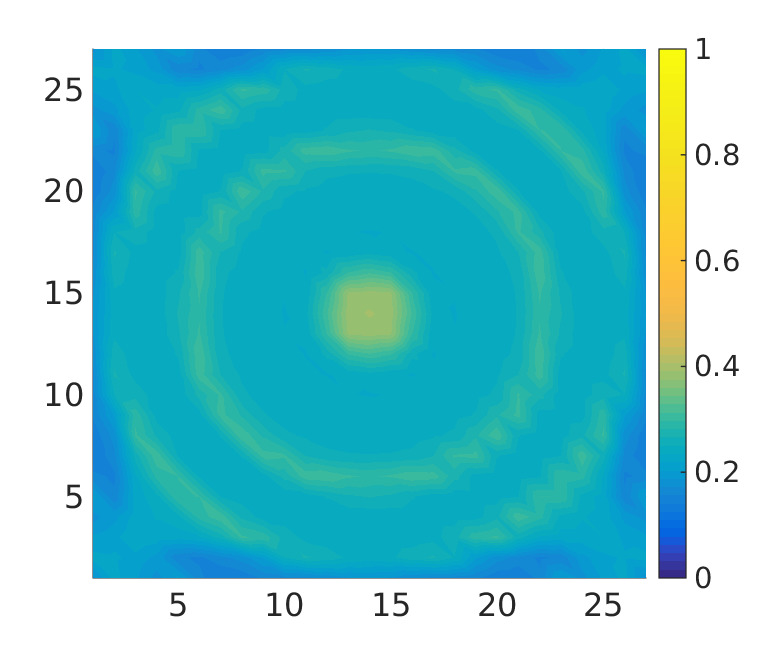}}
\subfigure[Second garbage  density, $t=5.4$]{\includegraphics[width=0.46\textwidth]{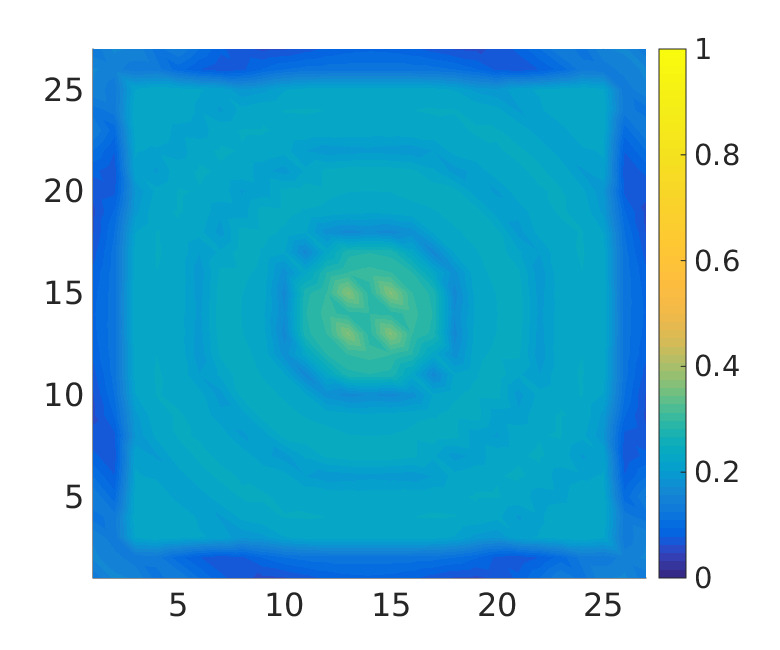}}
\end{center}
\vspace*{-0.5cm}\caption{On the left bacterial density over the lattice at $t=1.7, 3.6, 5.4$. On the right the second garbage density over the lattice at the same time. }
\label{fig:SG27_T5_b}
\end{figure}

Time evolutions of the mean values of the densities of the various populations together with their standard deviations over the lattice are shown in Figure \ref{fig:SG27_mean}, where we let the system evolve for 10 iterations, fixing $\tau=1$ and $L=27$, and where the subregion $\mathcal{O}$ consists in the two external rows/columns on the top-bottom and left-right of the lattice.  The plots show that the reintroduction of new nutrients in the system improves the overall density growth of all the populations in the lattice. This can be deduced since the mean values of the densities is here larger than in the case of absence of rules. In Fig.~\ref{fig:SG27_T5_b} we show the densities of the bacteria and of the second garbage over all the lattice at times $t=1.7, 3.6, 5.4$; we notice  the presence of a gradient (more evident for the garbage) in the external part of the lattice.

We end this section by observing that the rule considered here has a simple biological interpretation:  removing the \emph{hard} garbage along the borders of the lattice, and introducing  new nutrients,  corresponds to a periodic cleaning of the system accompanied by a supplementary source of nutrients allowing the long term sustainability of the ecosystem.
This explains why the asymptotic mean densities over the lattice of the bacteria and of the two garbages are larger than the mean densities obtained for the problem without rules.

\subsection{Rules acting on the parameters of the model}
\label{subsec::rule_ham}
Different rules can also be introduced to describe different biological effects, for instance changes in the metabolism of the bacteria due to stress factors. For instance, the increased density of the non recyclable garbage or the lack of nutrients are stress factors for the bacterial colonies, and cause changes in their behavior.

We start by using the values of the parameters and the initial conditions as stated above, and  take into account the dynamical mechanisms implied in each cell by the variation $D_{4,\alpha}$ of the scarcely recyclable garbage after any period of quantum evolution of the system of length $\tau$,
\begin{equation}
D_{4,\alpha}=n_{4,\alpha}(k \tau)-n_{4,\alpha}((k-1) \tau), \qquad k>1,\ \alpha=1,\ldots,L^2,
\end{equation}
to generate a new set of values for some of the parameters of the model according to the rule $\rho$ 
\begin{equation}\label{eq:ro}
\begin{array}{l}
\begin{cases}
\begin{array}{l}
\rho(\omega_{2,\alpha})=\omega_{2,\alpha}(1+10 D_{4,\alpha}),\\
\rho(\lambda_{2,\alpha})=\lambda_{2,\alpha}(1-10 D_{4,\alpha}),\\
\rho(\lambda_{4,\alpha})=\lambda_{4,\alpha}(1-10 D_{4,\alpha}),\\
\rho(\nu_{3,\alpha})=\nu_{3,\alpha}(1-10 D_{4,\alpha}).
\end{array}
\end{cases}
\end{array}
\end{equation}

Numerical simulations of the stepwise model describing the evolution of the system for consecutive time intervals in accordance with the Heisenberg representation and imposing repeatedly the rule $\rho$ have been compared to those obtained by considering the evolution according to the standard 
nonhomogeneous linear model.
The initial data are those used in the previous sections.

\begin{figure}[!]
\begin{center}
\subfigure[Bacterial densities at $t=0.5$]{\includegraphics[width=0.48\textwidth]{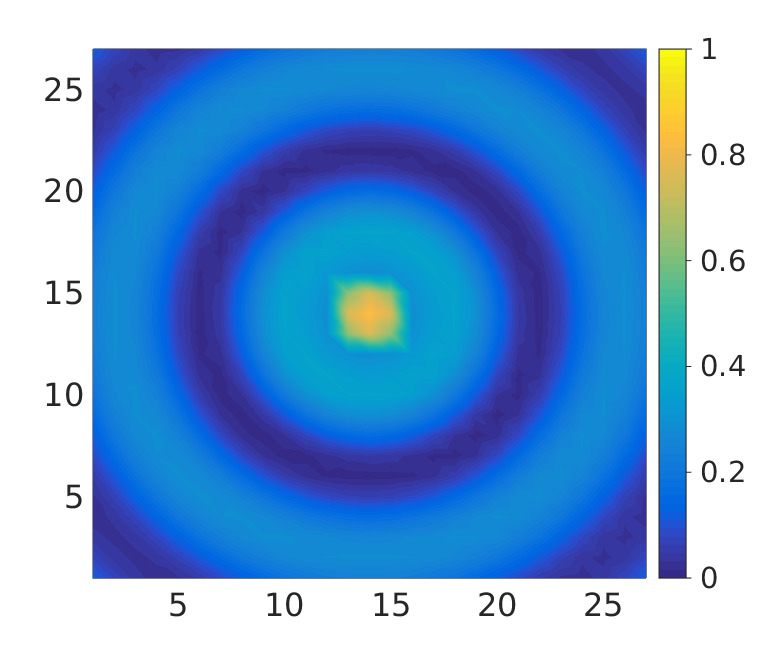}}
\subfigure[Bacterial densities at $t=1.7$]{\includegraphics[width=0.48\textwidth]{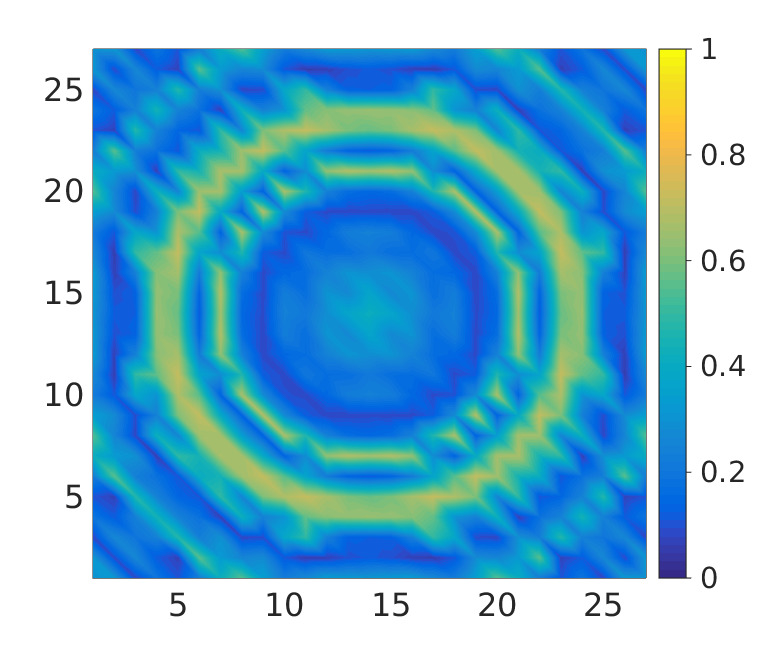}}\\
\subfigure[Bacterial densities at $t=3.6$]{\includegraphics[width=0.48\textwidth]{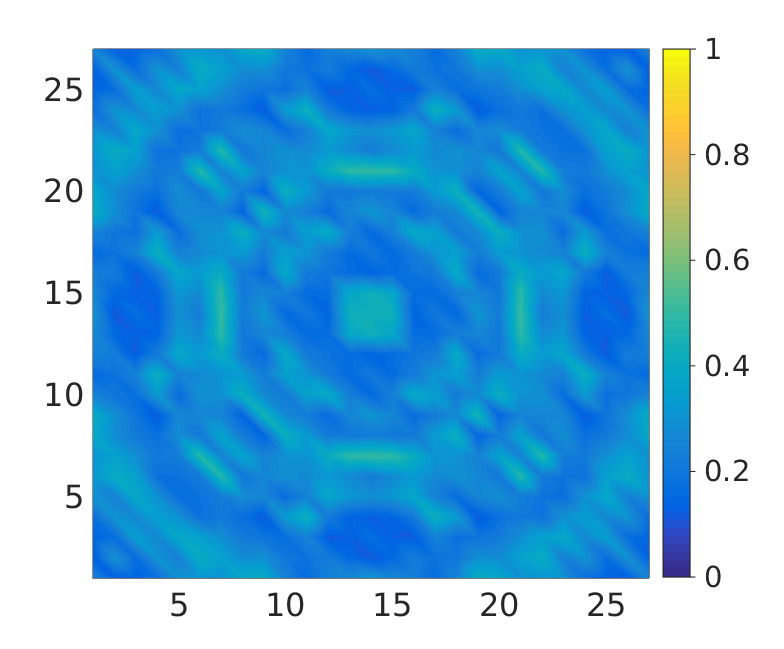}}
\subfigure[Bacterial densities at $t=5.4$]{\includegraphics[width=0.48\textwidth]{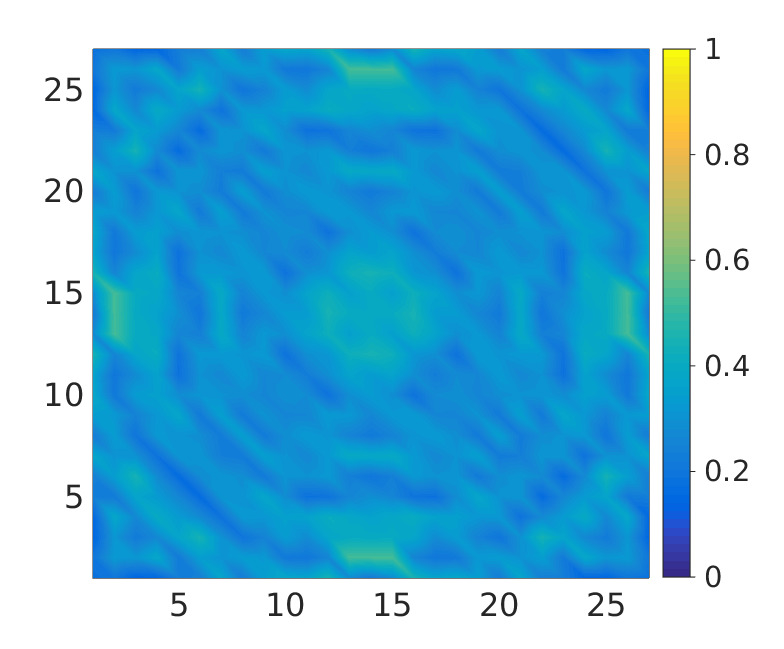}}\\
\end{center}
\caption{\label{fig:rules}Stepwise nonhomogeneous linear model with the rule $\rho$ and $\tau=1$. The frames show for each row the densities of the bacteria over the entire region at times
0.5, 1.7, 3.6, 5.4 respectively.}
\end{figure}

\begin{figure}[!]
\begin{center}
\subfigure{\includegraphics[width=0.48\textwidth]{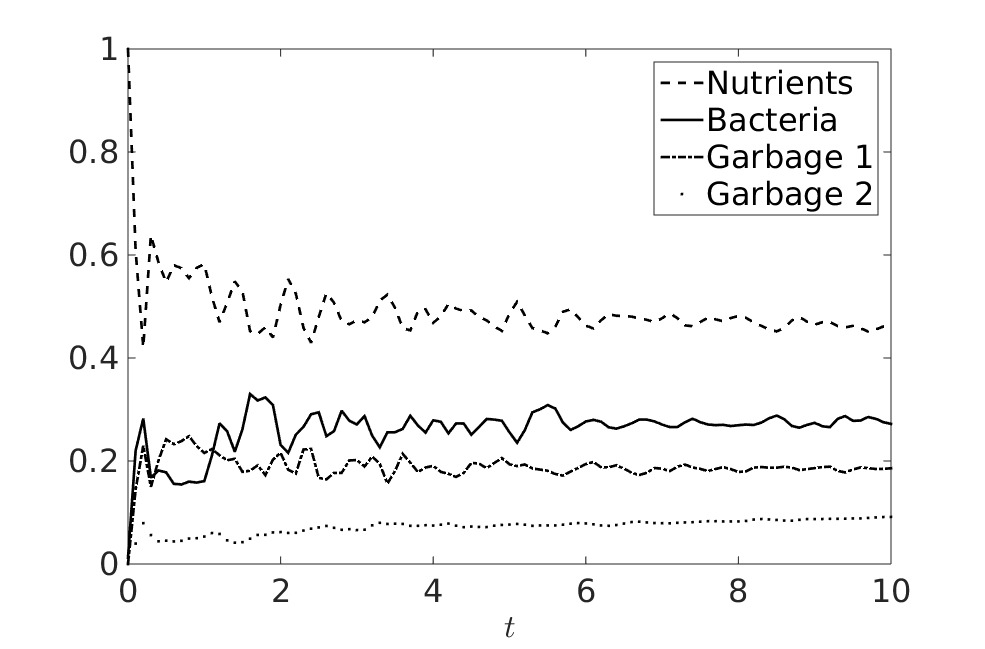}}
\subfigure{\includegraphics[width=0.48\textwidth]{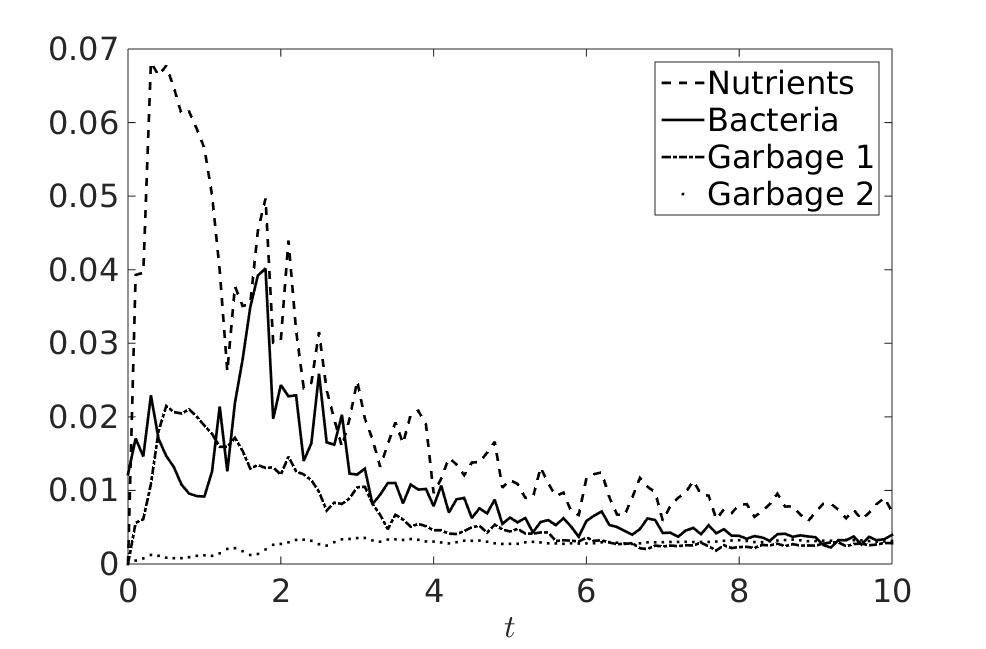}}
\end{center}
\caption{\label{fig:meanvarrules}Time evolution of the mean of the densities (on the left) and of the variance (on the right) of all the compartments of the ecosystem over a region described by means of the  stepwise nonhomogeneous linear model with the rule $\rho$ and $\tau=1$.}
\end{figure}

As visible in Fig.~\ref{fig:rules}, though the diffusion has the effect of distributing the bacteria all over the lattice (keep in mind that the radial inhomogeneity of the parameters determines the formation of symmetrical patterns), the validity of the proposed stepwise method, differently from what happens by adopting the standard linear model, is shown by the fact that the repeated application of the rule $\rho$ after several time steps improves the description of the behavior of bacterial populations in terms of long term survival. Therefore, the proposed stepwise spatial method naturally describes  the relevant changes of the metabolism of certain genera, which are able to survive for a very long time in terms of latent life when under negative stimuli, without external actions.
The mean values and the variances of the densities of all the compartments of the ecosystem exhibit  an oscillating behavior with decreasing variations and the tendency to stabilize, in the case of the mean values, and to become smaller and smaller, in the case of the variances (see Fig.~\ref{fig:meanvarrules} for the time evolutions corresponding to the simulation shown in Fig.~\ref{fig:rules}).

\section{Conclusions and perspectives}\label{sect5}

We have discussed how the notion of $(H,\rho)$-induced dynamics can be quite efficient to produce dynamical behaviors having some asymptotic non trivial limit, without the need of introducing any infinitely extended reservoir coupled to the system, or any explicit time dependent Hamiltonian. We have also shown that this is not really due to the explicit form of the rule $\rho$, but it is rather due to the presence itself of such a rule. In particular, we have compared different versions of the rule,  either changing the state of the system or modifying the parameters of the Hamiltonian of the system. Incidentally, this last choice could be seen as the use of a particularly simple time dependent Hamiltonian, which {modifies itself along the time evolution}. More precisely, it is more than this, as the results in the Appendix show. In fact, the application of the rule produces a piecewise constant quadratic Hamiltonian, whose related equations of motion can be solved quite easily, and then matched together. On the other hand, even for the simple example discussed in the Appendix, the choice of a time dependent Hamiltonian produces equations which, in general, are not so simple to solve, or can only be solved numerically. For this reason, our procedure appears more useful and technically much simpler. Moreover, as we have already discussed all along the paper, $\rho$ has also a natural interpretation in terms of \emph{measures performed over the physical system}, measures that, as it naturally happens in quantum mechanics, modify the system itself.

We should also mention that in the literature other ways are proposed to get some asymptotic behaviour for a given system. The approach based on master equation is one of this, \cite{rev3a,rev3b,Pkren2,rev3c}. This relies on the possibility of constructing a particularly simple dynamical equation, usually involving only few parameters. The model obtained in this way is easily solvable, but on the other hand several details of the interactions between the various agents of the original open system are lost. This is not what happens in our approach, which is much richer from this point of you. It is surely interesting, and is part of our future work, to compare in details these strategies. 

\appendix

\section*{Appendix 1:  A time dependent Hamiltonian}
\renewcommand{\theequation}{A.\arabic{equation}}

As we have seen, the concept of $(H,\rho)$-induced dynamics can be efficiently used in the description of systems which are expected to reach some equilibrium, during their time evolution. Now we want to see if this can also be achieved without using any rule, but assuming that $H$ can be explicitly dependent on time. To do this, we go back to the simple model considered in Section \ref{sect3}, and replace the Hamiltonian in (\ref{MM23}) by a similar, but time dependent, operator:
\begin{equation}
H(t)=\omega_1(t)a_1^\dagger a_1+\omega_2(t)a_2^\dagger a_2 +\lambda(t)(a_1^\dagger a_2+a_2^\dagger a_1). \label{31}
\end{equation}
Here
$\omega_j(t)$ and $\lambda(t)$ are real functions of the time. If we assume $\mathcal{S}$ is initially in the state $\Psi_0=\sum_{n_1,n_2=0}^1 c_{n_1,n_2}\varphi_{n_1,n_2}$, with $\sum_{n_1,n_2=0}^1 |c_{n_1,n_2}|^2=1$, the Schr\"odinger equation $i\dot\Psi(t)=H(t)\Psi(t)$ can be rewritten, in terms of the time evolution of the coefficients $c_{n_1,n_2}$, as follows:
\begin{equation}
\begin{aligned}
&i \dot c_{0,0}(t)=0,\\  
&i \dot c_{1,0}(t)=\omega_1(t) c_{1,0}(t) +\lambda(t)c_{0,1}(t),\\
&i \dot c_{0,1}(t)=\lambda(t) c_{1,0}(t) +\omega_2(t)c_{0,1}(t),\\
&i \dot c_{1,1}(t)=\left(\omega_1(t)  +\omega_2(t)\right)c_{1,1}(t),
\end{aligned}
\label{NA16}
\end{equation}
with $c_{n_1,n_2}(0)=c_{n_1,n_2}$. It is clear that
\begin{equation}
c_{0,0}(t)=c_{0,0}, \qquad c_{1,1}(t)=c_{1,1} \exp(-i(\Omega_1(t)+\Omega_2(t))),
\label{NA17}
\end{equation}
where $\Omega_j(t)=\int_0^t\omega_j(t_1)dt_1$, so that $|c_{0,0}(t)|^2=|c_{0,0}|^2$ and $|c_{1,1}(t)|^2=|c_{1,1}|^2$, for all $t$. Of course, the equations in (\ref{NA16}) suggest that the time evolution of $c_{0,1}(t)$ and $c_{1,0}(t)$ is decoupled from $c_{0,0}(t)$ and $c_{1,1}(t)$. However, this is not really so. In fact, since the time evolution of $\mathcal{S}$ preserves the norm of $\Psi(t)$, $\|\Psi(t)\|=\|\Psi(0)\|$, we find that the coefficients $c_{n_1,n_2}(t)$ must satisfy the constraint $
\sum_{n_1,n_2=0}^1 |c_{n_1,n_2}(t)|^2=1$, which implies that
\begin{equation}
|c_{0,1}(t)|^2+|c_{1,0}(t)|^2=1-|c_{0,0}(t)|^2-|c_{1,1}(t)|^2=1-|c_{0,0}|^2-|c_{1,1}|^2,
\label{NA18}
\end{equation}
for all $t$. Extending what we have done in subsection \ref{sect3a}, where we have computed $n_j(t)=\left<\varphi_{n_1,n_2},\right.\allowbreak\left.
\hat n_j(t)\varphi_{n_1,n_2}\right>$, we are interested here in computing $N_j^\Psi(t)=\left<\Psi(t),
\hat n_j\Psi(t)\right>$. It turns out that
\begin{equation}
N_1^\Psi(t)=|c_{1,0}(t)|^2+|c_{1,1}|^2, \qquad N_2^\Psi(t)=|c_{0,1}(t)|^2+|c_{1,1}|^2,
\label{NA20}
\end{equation}
which shows, in particular, that $c_{0,0}$ does not enter this computation, and that we do not need to follow the time evolution of $c_{1,1}(t)$, since we only need its square modulus, which is constant. Concerning $c_{1,0}(t)$ and $c_{0,1}(t)$, if we introduce
\begin{equation}
d_{1,0}(t)=c_{1,0}(t)e^{i\Omega_1(t)}, \qquad d_{0,1}(t)=c_{0,1}(t)e^{i\Omega_2(t)},
\label{NA21}
\end{equation}
as well as $\Phi(t)=\Omega_1(t)-\Omega_2(t)$, we get the following system of coupled differential equations,
\begin{equation}
i \dot d_{1,0}(t)= \lambda(t)e^{i\Phi(t)}d_{0,1}(t),\qquad
i \dot d_{0,1}(t)=\lambda(t)e^{-i\Phi(t)}d_{1,0}(t),
\label{NA22} 
\end{equation}
with initial conditions $d_{1,0}(0)=c_{1,0}$ and $d_{0,1}(0)=c_{0,1}$. Moreover $N_1^\Psi(t)=|d_{1,0}(t)|^2+|c_{1,1}|^2$ and  $N_2^\Psi(t)=|d_{0,1}(t)|^2+|c_{1,1}|^2$. It is clear that, if $|c_{0,0}|^2+|c_{1,1}|^2=1$, then both $N_1^\Psi(t)$ and $N_2^\Psi(t)$ admit an asymptotic value, which is $N_j^\Psi(\infty):=\lim_{t\rightarrow\infty}N_j^\Psi(t)=|c_{1,1}|^2$, $j=1,2$. This is a simple consequence of (\ref{NA18}), which implies that, in this case, $d_{1,0}(t)=d_{0,1}(t)=0$, for all $t$. From now on we will assume that $c_{0,1}$ or $c_{1,0}$, or both, are different from zero. Then, since
\begin{equation}
\frac{d|d_{1,0}(t)|^2}{dt}=2i\lambda(t)\Im\left[d_{1,0}(t)\overline{d_{0,1}(t)}e^{-i\Phi(t)}\right],
\end{equation}
it is clear that $|d_{1,0}(t)|$ converges to some asymptotic value (and $N_1^\Psi(t)$ and $N_2^\Psi(t)$ as well, as a  consequence), if either (i) $\lambda(t)$ converges to zero when $t$ diverges, or when (ii) $\lim_{t\rightarrow\infty}\Im\left[d_{1,0}(t)\overline{d_{0,1}(t)}e^{-i\Phi(t)}\right]=0$. The case (i) is completely clear: in absence of interaction the two densities have no way (and no reason) to change in time. On the other hand, the case (ii) is not particularly useful, since to check the condition we should solve the dynamical problem and compute  $d_{1,0}(t)$ and $d_{0,1}(t)$ first.

In order to find more cases which are "under control", we rewrite (\ref{NA22}) as a second order equation. Assuming that $\lambda(t)$ is never zero, we find that
\begin{equation}
\ddot d_{1,0}(t)-\frac{\dot\lambda(t)+i\lambda(t)(\omega_1(t)-\omega_2(t))}{\lambda(t)}\dot d_{1,0}(t)+\lambda^2(t) d_{1,0}(t)=0,
\label{NA23}
\end{equation}
with initial conditions $d_{1,0}(0)=c_{1,0}$ and $\dot d_{1,0}(0)=-i\lambda(0)c_{0,1}$. The equation for $d_{0,1}(t)$ is simply the complex conjugate of equation (\ref{NA23}).

\begin{figure}
\begin{center}
\subfigure[]{\includegraphics[width=0.45\textwidth]{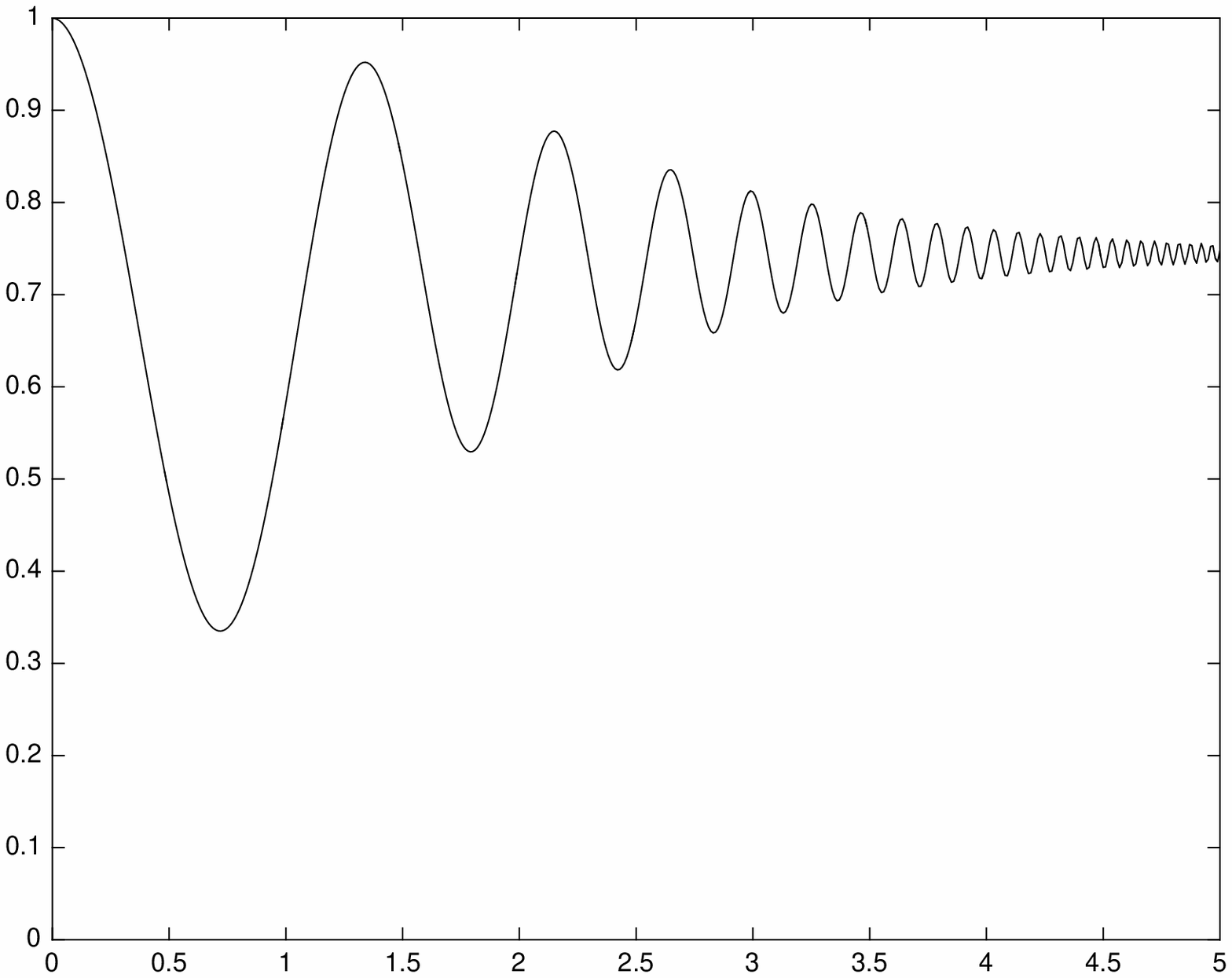}}
\subfigure[]{\includegraphics[width=0.45\textwidth]{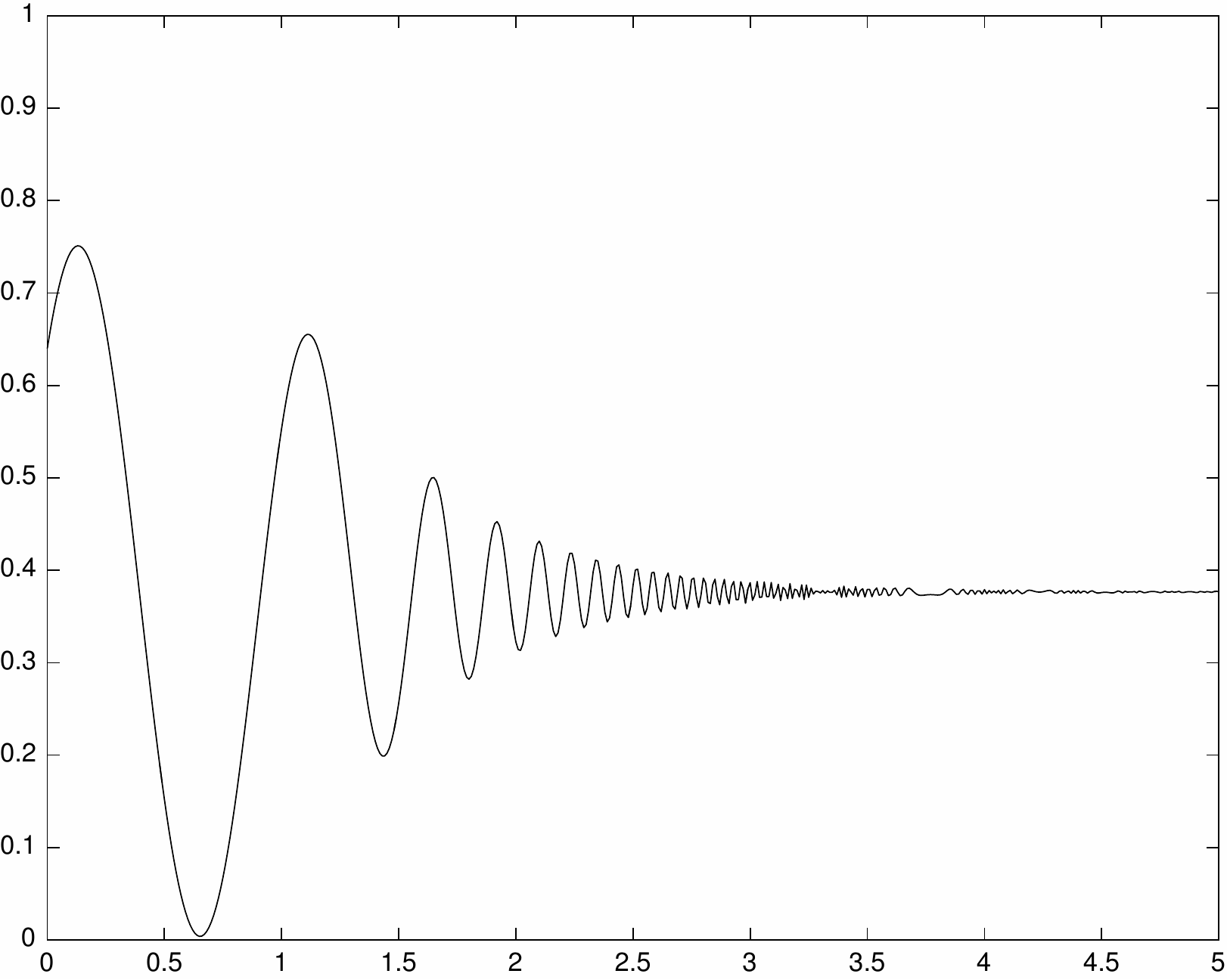}}
\end{center}
\caption{\label{figadd}$|d_{1,0}(t)|^2$ for $\omega_1(t)-\omega_2(t)=-(\exp(t)+1)$, $\lambda(t)=\sqrt{3}$, $d_{1,0}(0)=1$ and $\dot d_{1,0}(0)=0$ (left); $|d_{1,0}(t)|^2$ for $\omega_1(t)-\omega_2(t)=-t^2(\exp(t)+1)$, $\lambda(t)=3$, $d_{1,0}(0)=.8$ and $\dot d_{1,0}(0)=1$ (right).}
\end{figure}

It is easy to find situations in which neither $d_{1,0}(t)$ nor its modulus admits an asymptotic value. For instance, if we fix $\lambda(t)$ in such a way $\dot\lambda(t)+i\lambda(t)(\omega_1(t)-\omega_2(t))=0$, and we further restrict to $\omega_1(t)=\omega_2(t)$, then $d_{1,0}(t)$ is an oscillating function. The same conclusion is deduced if we try to keep in the equation (\ref{NA23}) the term proportional to $\dot d_{1,0}(t)$, to have a sort of friction in the system. However, the only solution consistent with the fact that $H(t)=H^\dagger(t)$ is again an oscillating function. However, this does not imply that no solution $d_{1,0}(t)$ exists such that $|d_{1,0}(t)|$ admits an asymptotic value. This happens, in fact, if $\lambda(t)=\lambda_0$ is constant and $|\omega_1(t)-\omega_2(t)|$ is a certain increasing function. For instance, in Figure \ref{figadd}(a) we plot $|d_{1,0}(t)|^2$ for $\omega_1(t)-\omega_2(t)=-(\exp(t)+1)$ and $\lambda(t)=\sqrt{3}$. We are fixing here the initial conditions $d_{1,0}(0)=1$ and $\dot d_{1,0}(0)=0$. We see that, indeed, an asymptotic value is reached for $|d_{1,0}(t)|^2$, which means that also $|d_{0,1}(t)|^2$ converges to a limiting values when $t$ diverges.

The same conclusion can be deduced considering Figure \ref{figadd}(b), where we plot again $|d_{1,0}(t)|^2$ for $\omega_1(t)-\omega_2(t)=-t^2(\exp(t)+1)$ and $\lambda(t)=3$. The initial conditions are here $d_{1,0}(0)=.8$ and $\dot d_{1,0}(0)=1$.

The plots we get here are qualitatively similar to those depicted in Fig.~\ref{fig:2mode_ruleAB}.
The conclusion of this analysis is that, as we have claimed several times along the paper, the use of the $(H,\rho)$--induced dynamics is not extremely different  from adopting a time dependent Hamiltonian.
In fact, when using the rule, the parameters of the Hamiltonian become step-wise constant functions. Moreover, parameters adjust themselves in a natural way leading to an equilibrium. On the other hand,   Hamiltonian $H(t)$ in (\ref{31}) requires a very \emph{ad hoc} choice of the functions to produce such an equilibrium. Last but not least, the use of the rule is simpler from a technical point of view.

\section*{Acknowledgements}
The authors acknowledge partial support from GNFM of the INdAM. 
F.~B. and F.~G. acknowledge partial support from Palermo University. F.B. thanks the 
\emph{Distinguished Visitor Program} of the Faculty of Science of the University of Cape Town, 2017.

\end{document}